\documentclass{aastex62}
\usepackage{amsmath,amssymb}
\usepackage{gensymb}
\usepackage{appendix}
\usepackage{physics}
\usepackage{multirow}
\usepackage{textcomp}
\usepackage{array}
\usepackage{threeparttable}
\newcolumntype{P}[1]{>{\centering\arraybackslash}p{#1}}

\hypersetup{linkcolor=red,citecolor=purple,filecolor=cyan,urlcolor=magenta}

\graphicspath{{./}{Figures/}}

\received{}
\revised{}
\accepted{}

\shorttitle{A Non-Hydrostatic Atmospheric Model with the VIC scheme}
\shortauthors{Ge et al.}

\begin{document}

\title{A Global Non-Hydrostatic Atmospheric Model with a Mass and Energy Conserving Vertically-Implicit-Correction (VIC) Scheme}

\correspondingauthor{Huazhi Ge}
\email{huazhige@ucsc.edu}

\author[0000-0001-6719-0759]{Huazhi Ge}
\affil{Department of Earth and Planetary Sciences, 
The University of California, Santa Cruz, 1156 High Street, Santa Cruz, CA 95064, USA}

\author{Cheng Li}
\affiliation{Astronomy Department, The University of California, Berkeley, CA 94720, USA}

\author{Xi Zhang}
\affiliation{Department of Earth and Planetary Sciences, 
The University of California, Santa Cruz, 1156 High Street, Santa Cruz, CA 95064, USA}

\author{Dongwook Lee}
\affiliation{Applied Mathematics, 
The University of California, Santa Cruz, 1156 High Street, Santa Cruz, CA 95064, USA}



\begin{abstract}

Global non-hydrostatic atmospheric models are becoming increasingly important for studying the climates of planets and exoplanets. However, such models suffer from computational difficulties due to the large aspect ratio between the horizontal and vertical directions. To overcome this problem, we developed a global model using a vertically-implicit-correction (VIC) scheme in which the integration time step is no longer limited by the propagation of acoustic waves in the vertical. We proved that our model, based on the $\rm Athena^{++}$ framework and its extension for planetary atmospheres - SNAP (Simulating Non-hydrostatic Atmosphere on Planets), rigorously conserves mass and energy in finite volume simulations. We found that traditional numerical stabilizers such as hyper-viscosity and divergence damping are not needed when using the VIC scheme, which greatly simplifies the numerical implementation and improves stability. We present 
simulation results ranging from 1D linear waves to 3D global circulations with and without the VIC scheme. These tests demonstrate that our formulation correctly tracks local turbulent motions, produces Kelvin-Helmholtz instability, and generates a super-rotating jet on hot Jupiters. Employing this VIC scheme improves the computational efficiency of global simulations by more than two orders of magnitude compared to an explicit model and facilitates the capability of simulating a wide range of planetary atmospheres both regionally and globally.

\end{abstract}




\section{Introduction} \label{sec:intro}

GCMs (General Circulation Models) are numerical tools for studying the weather and climate of planetary atmospheres. They solve the hydrodynamic equations on a sphere including a whole range of additional physical processes such as rotation, radiation, and tracer transport. To improve the computational efficiency, various levels of assumptions can be adopted to simplify the calculation. Some famous forms include the quasi-geostrophic equations, shallow water equations, primitive equations, Boussinesq equations, and anelastic equations \citep[e.g.,][]{holton2004introduction,pedlosky2013geophysical,vallis2017atmospheric,holton2016dynamic}. Comparisons of some forms of these equations are detailed in \cite{white2005consistent}, \cite{mayne2014unified}, and \cite{mayne2019limits}. Most GCMs adopt the primitive equations to study the general circulation of planetary atmospheres by assuming a ``thin shell" atmosphere in hydrostatic equilibrium and neglecting some terms in the momentum equations (the ``traditional approximations")  \citep{boer1984canadian,held1994proposal,dowling1998explicit,adcroft2004implementation,holton2004introduction}. Though this approach has been successful in exploring global features of the Earth's atmosphere for decades, the usage of primitive equations has some limitations when applied to diverse planetary atmospheres other than that of Earth. For example, \cite{mayne2019limits} showed the simulations of tidally locked sub-Neptune atmospheres using the ``full" dynamical equations without the above approximations are different from those derived using the primitive equations, with the differences attributed to the traditional approximation (also see \citealt{tokano2013wind} for 
a study on slow rotators like Venus and Titan). Therefore the full set of Euler equations has to be applied to study weather and climate in the atmospheric regime where conventional assumptions break down (see Appendix~\ref{sec:governing-equ} for the detailed formulation).

In addition, planetary atmospheres broadly exhibit diverse behaviors associated with complex physical processes, such as atmospheric collapse, surface interactions, multi-layer moist convection, and interactions with magnetic fields. These processes are usually parameterized in conventional GCMs for simplification and computational efficiency \citep[e.g.,][]{suarez1983parameterization,newman2002modeling,schneider2009formation,lian2010generation}. The input parameters in these designed schemes are often adjusted to match observations. For example, convection parameterizations are commonly adopted in Earth atmosphere simulations, such as quasi-equilibrium schemes \citep[e.g.,][]{betts1986new,emanuel1991scheme,emanuel1993effect}. It has recently been shown that applying convection parameterizations developed for the Earth to study tidally locked terrestrial planets might lead to overestimation of the heat redistribution efficiency between hemispheres compared with the high-resolution convection-resolving simulations. In the latter, the full set of Euler equations is also needed \citep{sergeev2020atmospheric}. 

Non-hydrostatic GCMs solving the full set of Euler equations emerged at the turn of this century. However, directly solving the Euler equations is usually computationally expensive for global atmospheric simulations due to the numerical limitation imposed by the meteorologically-trivial but fast-propagating acoustic waves as well as by the large aspect ratio between the vertical and the horizontal directions. The difficulty is more severe when simulating atmospheric circulations on cold and large bodies, such as gas giants and ice giants. For example, at the 1 bar pressure level, Jupiter's radiative cooling timescale is about 10 times slower than the Earth's, meaning it requires a much longer time to integrate to a steady and fully-evolved state. Solving the full set of Euler equations with a vertically-implicit scheme allows the use of a large time step by damping the vertically-propagating acoustic waves in the vertical direction. Several non-hydrostatic models with a variety of vertically-implicit schemes have been developed to study the Earth atmosphere (see \citealt{ullrich2017dcmip2016} for a recent model intercomparison) and exoplanetary atmospheres \citep{mayne2014unified,mendoncca2016thor,deitrick2019thor}.

It is also important to develop a GCM dynamical core that can rigorously conserve total mass and energy in a closed domain. The conservation of total axial angular momentum (AAM) is also crucial for studying atmospheric dynamics, especially the zonal jet patterns. Satisfying these conservation laws is particularly important for simulations that require very long time integration of more than a decade or even centuries, such as that on Venus (slow rotator) \citep{lebonnois2013models,mendoncca2020modelling}, giant planets (cold atmospheres) \citep{schneider2009formation,liu2010mechanisms,spiga2020global}, and hot Jupiters (long radiative timescale in the deep atmosphere) \citep{mayne2017results,deitrick2019thor,wang2020extremely}. In these cases, the numerical schemes used in the dynamical cores have to be carefully designed with the capability of satisfying conservation laws. Any continuous loss or increase of mass, energy, or AAM prohibits the numerical model to reach a steady state. However, previously presented vertically-implicit schemes seldom provide detailed proof of such conservation.

Here we present a three-dimensional (3D) non-hydrostatic GCM with a state-of-the-art mass and energy conserving Vertically-Implicit-Correction (VIC) scheme based on the $\rm Athena^{++}$ framework \citep{stone2020athena++} and its extension for planetary atmospheres, SNAP (Simulating Non-hydrostatic Atmospheres on Planets) \citep{li2019simulating}. In Section~\ref{sec:num-HEVI}, we describe the governing equations and the algorithm of our VIC scheme including a modified time integration scheme and a dimensionally-unsplit method. In Section~\ref{sec:conservation-law}, we prove that our VIC scheme satisfies conservation laws. In Section~\ref{sec:benchmark}, we present numerical solutions of local simulations (e.g., linear wave test, Straka sinking bubble test, and Robert rising bubble test) and global simulations (e.g., Held-Suarez test and shallow hot Jupiter benchmark test). Local simulations are designed to validate the damping of acoustic waves and the capability of resolving turbulence. The global simulations are designed to validate the model performance such as mass, energy, and AAM conservation and jet formation. Finally, we summarize our works and list our future plans in Section~\ref{sec:conclusion}.


\section{Numerical Scheme for Implicit SNAP} \label{sec:num-HEVI}

We use the Finite Volume Method (FVM) to discretize the Euler equations. Conservative variables -- density ($\rho$), momentum in three directions ($\rho u$, $\rho v$, and $\rho w$), and total energy ($E$), $\boldsymbol{Q}=(\rho, \rho u, \rho v, \rho w, E)^T$ -- are solved in the unstaggered control volumes in each time step. Total energy is the internal energy plus the kinetic energy, $E = p/(\gamma-1) + \rho (u^2 + v^2 + w^2)/2$. In a specific cell, we can write the integrated form of Euler equations as,

\begin{equation}
      \label{equ:integreted-Euler-simple}
      \iiint_V \frac{\partial}{\partial t}\boldsymbol{Q}(\boldsymbol{x},t) \, dV + \iiint_V \nabla \cdot \boldsymbol{f}(\boldsymbol{Q}) dV = 
      \iiint_V \boldsymbol{\mathcal{F}}(\boldsymbol{Q},\boldsymbol{x},t) dV,
\end{equation}
where $\boldsymbol{Q}=(\rho,\rho u,\rho v, \rho w, E)^T$ is the vector of conservative variables; $\boldsymbol{f}(\boldsymbol{Q})$ are fluxes of $\boldsymbol{Q}$ in three directions, which are $\boldsymbol{F} = (\rho u, \rho uu+p, \rho uv, \rho uw, E+p)^{T}$, $\boldsymbol{G} = (\rho v, \rho vu, \rho vv+P, \rho vw, E+p)^{T}$, and $\boldsymbol{H} = (\rho w, \rho wu, \rho wv, \rho ww+p, E+p)^{T}$; $\boldsymbol{\mathcal{F}}(\boldsymbol{Q},\boldsymbol{x},t)$ is the vector of body forces (i.e., source terms), such as gravity. The detailed Euler equations in different coordinate systems are presented in Appendix~\ref{sec:governing-equ}.

\subsection{VIC in the Forward-Euler Time Integration Scheme} \label{sec:implicit-corr}

We start with describing our model Euler equations in 1D to depict the derivation and formulation of our VIC scheme. Discretized using the simple forward-Euler time integration scheme, the governing equation can be given by,

\begin{equation}
    \label{equ:1d-euler}
    \frac{\boldsymbol{Q}_{i}^{n+1} - \boldsymbol{Q}_{i}^{n}}{\Delta t} + \frac{\sigma_{i+1/2}\boldsymbol{F}_{i+1/2}^{n+1} - \sigma_{i-1/2}\boldsymbol{F}_{i-1/2}^{n+1}}{\Delta V_{i}} = \boldsymbol{X}^{n+1}_{i} + \boldsymbol{Y}^{n}_{i},
\end{equation}
where $\boldsymbol{Q}_{i}^{n}$ represents the vector of volume-averaged conservative variables in the $i$-th cell at the $n$-th time step (i.e., current time step); $\boldsymbol{F}_{i-1/2}^{n+1}$ and $\boldsymbol{F}_{i+1/2}^{n+1}$ are numerical fluxes across the left (bottom) and the right (top) interfaces of the $i$-th cell at the $(n+1)$-th time step (i.e., next time step), respectively; $\boldsymbol{Y}^{n}_{i}$ and $\boldsymbol{X}^{n+1}_{i}$ are explicitly and implicitly treated source terms, respectively. $\Delta V_{i}$ is the volume of the cell, which is constant in the Cartesian coordinate system but changes with latitude and radius in the spherical-polar coordinate system. $\sigma_{i+1/2}$ and $\sigma_{i-1/2}$ are face areas of the $i$-th cell in the vertical direction. On the right-hand side of the equation, most of the source terms, such as gravitational acceleration, Coriolis force, and centrifugal force, can be treated implicitly (i.e., included in $\boldsymbol{X}^{n+1}_{i}$) because they 
can be written analytically in terms of flux Jacobian, which is the first-order partial derivative of the forcing with respect to $\boldsymbol{Q}$. Radiative forcing are treated explicitly because it usually does not have an analytical Jacobian.

Once we know $\Delta \boldsymbol{Q}_i = \boldsymbol{Q}_{i}^{n+1} - \boldsymbol{Q}_{i}^{n}$, then we can update $\boldsymbol{Q}_{i}^{n+1}$  by $\boldsymbol{Q}_{i}^{n+1} = \Delta \boldsymbol{Q}_i + \boldsymbol{Q}_{i}^{n}$. Then, the problem becomes how to acquire the flux and source terms at $(n+1)$-th time step, $\boldsymbol{F}_{i-1/2}^{n+1}$, $\boldsymbol{F}_{i+1/2}^{n+1}$, and $\boldsymbol{X}^{n+1}_{i}$, from the conservative variables at the $n$-th step. Following the ideas in \cite{fernandez1988implicit} and \cite{viozat1997implicit}, we use the Roe scheme to acquire the implicit fluxes across cell interfaces at the $(n+1)$-th time step \citep{roe1981approximate}, 

\begin{equation}
    \label{equ:Roe-n+1}
    \boldsymbol{F}_{i+1/2}^{n+1} = \frac{1}{2} ( \boldsymbol{F}_{i}^{n+1} + \boldsymbol{F}_{i+1}^{n+1} - \vert A_{i+1/2}^{n+1} \vert (\boldsymbol{Q}_{i+1}^{n+1} - \boldsymbol{Q}_{i}^{n+1}) ),
\end{equation}
where $\vert A_{i+1/2}^{n+1} \vert$ is the Roe-averaged flux Jacobian across the right interface of the $i$-th cell \citep{roe1981approximate}. This formulation can also be applied to the fluxes at the $n$-th time level. $\vert A_{i+1/2}^{n} \vert$ performs as a ``stabilization term'' to stabilize the flux at the cell interface. It is different from the hyper-viscosity which is implemented in many GCMs. It is commonly considered as a stabilization term in the Roe scheme. The detailed discussion of $\vert A_{i+1/2}^{n} \vert$ is well presented in \citet{roe1981approximate}, \citet[Chap. 4.14]{leveque2002finite}, and \citet[Chap. 11]{toro2013riemann}. We provide the analytical derivation of $\vert A_{i+1/2}^{n} \vert$ in Appendix~\ref{sec:explain-A}. 

Note that the Roe-averaged Jacobian at the $(n+1)$-th time step, $\vert A_{i+1/2}^{n+1}\vert$, is unknown at the $n$-th time step. As a workaround, we consider the first-order approximation, $\vert A_{i+1/2}^{n} \vert = \vert A_{i+1/2}^{n+1}\vert + \mathcal{O}(\Delta t)$, which can be justified by two reasons: (1) because $\vert A \vert$ is dictated by the flow and acoustic speed and they are locally uniform in the atmosphere, therefore the temporal evolution of $\vert A \vert$ remains trivial over a small time step $\Delta t$ ($\Delta t $ is usually much smaller than $10^3$ seconds); (2) the desired numerical stability of the Roe scheme is not severely compromised by the approximation of using $\vert A_{i+1/2}^{n} \vert$, since $\vert A \vert$ performs as the numerical diffusion term. We will demonstrate the validity of this approximation by benchmark simulations in Section~\ref{sec:benchmark}. 

To simplify the notation in this section, we define the flux Jacobian, $\boldsymbol{J}_{i}^{n}$, and the Jacobian of the source term, $\boldsymbol{X'}_{i}^{n}$, as,

\begin{equation}
    \label{equ:define-flux-Jacobian}
    \boldsymbol{J}_{i}^{n} \equiv \frac{\partial \boldsymbol{F}}{\partial \boldsymbol{Q}}\Big\vert_{i}^{n}
\end{equation}
and
\begin{equation}
    \label{equ:define-forcing-Jacobian}
    \boldsymbol{X'}_{i}^{n} \equiv \frac{\partial {\boldsymbol{X}}}{\partial \boldsymbol{Q}}\Big\vert_{i}^{n},
\end{equation}

The numerical flux, $\boldsymbol{F}_{i}^{n+1}$, and the implicitly treated source terms, $\boldsymbol{X}^{n+1}_{i}$, at the $(n+1)$-th time step are unknown. To compute these values, we use a Taylor expansion to linearize the flux at the $(n+1)$-th time step. For example, we predict $\boldsymbol{F}_{i}^{n+1}$ and $\boldsymbol{X}^{n+1}_{i}$, at the $(n+1)$-th step from the $n$-th time step by a second-order accurate Taylor expansion,

\begin{equation}
    \label{equ:flux-taylor}
    \boldsymbol{F}_{i}^{n+1} \approx \boldsymbol{F}_{i}^{n} + \boldsymbol{J}_{i}^{n} \Delta \boldsymbol{Q}_{i} + \mathcal{O}[(\Delta {\boldsymbol{Q}}_{i})^2]
\end{equation}
and
\begin{equation}
    \label{equ:source-taylor}
    \boldsymbol{X}^{n+1}_{i} \approx {\boldsymbol{X}}_{i}^{n} + \boldsymbol{X'}_{i}^{n}\Delta \boldsymbol{Q}_{i} + \mathcal{O}[(\Delta {\boldsymbol{Q}_{i})^2}],
\end{equation}
where $\boldsymbol{F}_{i}^{n}$ and $\boldsymbol{X}^{n}_{i}$ are numerical flux and source terms at $n$-th time step, respectively, and $\Delta \boldsymbol{Q}_{i} = \boldsymbol{Q}^{n+1}_{i} - \boldsymbol{Q}^{n}_{i}$. As previously mentioned, our numerical scheme solves $\Delta \boldsymbol{Q}_{i}$ to obtain $\boldsymbol{Q}^{n+1}_{i}$.

Note that Roe flux is not available at cell centers (i.e., $\boldsymbol{F}_{i}^{n}$ and $\boldsymbol{F}_{i+1}^{n}$ in Equation~(\ref{equ:flux-taylor})) because fluxes are evaluated at the cell interfaces. These terms can be eliminated by Equation~(\ref{equ:Roe-n+1}) at $n$-th time step to assemble the interface flux at time step $n$ (i.e., $\boldsymbol{F}_{i+1/2}^{n}$),

\begin{equation}
    \label{equ:Roe-n}
    \boldsymbol{F}_{i}^{n} + \boldsymbol{F}_{i+1}^{n} = 2\boldsymbol{F}_{i+1/2}^{n} + \vert A_{i+1/2}^{n} \vert (\boldsymbol{Q}_{i+1}^{n} - \boldsymbol{Q}_{i}^{n}) .
\end{equation}

Then, we can assemble linear equations about $\Delta \boldsymbol{Q}_{i}$ by combining Equations~(\ref{equ:flux-taylor}), (\ref{equ:source-taylor}), and (\ref{equ:Roe-n}). As a consequence, the tridiagonal linear system can be rewritten as,

\begin{equation}
\label{equ:linear_sys}
        \frac{1}{\Delta V_{i} }\Big[ c_{i} \Delta \boldsymbol{Q}_{i-1} +
        (\frac{\boldsymbol{\mathcal{I}}}{\Delta t} - \boldsymbol{X'}_{i}^{n} + a_{i})\Delta \boldsymbol{Q}_{i} + 
        b_{i}\Delta \boldsymbol{Q}_{i+1} \Big]= 
        - \frac{\Delta \sigma_{i+1/2}\boldsymbol{F}_{i+1/2}^{n} - \Delta \sigma_{i-1/2}\boldsymbol{F}_{i-1/2}^{n}}{\Delta V_{i} }+ 
        \boldsymbol{X}^{n}_{i} + {\boldsymbol{Y}^{n}_{i}},
\end{equation}
with coefficients $a_i$, $b_i$, and $c_i$ being,
\begin{equation}
\label{equ:a}
    a_{i} = \frac{1}{2} \Big[ (\sigma_{i+1/2} - \sigma_{i-1/2})\boldsymbol{J}_{i}^{n} +
    \sigma_{i-1/2}\vert A_{i-1/2}^{n} \vert + \sigma_{i+1/2}\vert A_{i+1/2}^{n} \vert \Big],
\end{equation}

\begin{equation}
\label{equ:b}
    b_{i} = \frac{\sigma_{i+1/2}}{2} \Big( \boldsymbol{J}_{i+1}^{n}- \vert A_{i+1/2}^{n} \vert \Big),
\end{equation}

and

\begin{equation}
\label{equ:c}
    c_{i} = -\frac{\sigma_{i-1/2}}{2} \Big( \boldsymbol{J}_{i-1}^{n} + \vert A_{i-1/2}^{n} \vert \Big),
\end{equation}
where $\boldsymbol{\mathcal{I}}$ is the identity matrix. When $i = 1$ and $i = m$, these equations shall be modified to satisfy different boundary conditions, which is detailed in Section~\ref{sec:conservation-law} and Appendix~\ref{sec:reflecting-prove}.

We use the fifth-order accurate Weighted Essentially Non-Oscillation (WENO5) reconstruction scheme and Riemann solvers in SNAP \citep{li2019simulating} to compute the Riemann state at each interface $i \pm 1/2$, which are used to solve the Riemann problems at $i \pm 1/2$ to calculate the Roe fluxes, $\boldsymbol{F}_{i+1/2}^{n}$ and $\boldsymbol{F}_{i-1/2}^{n}$. Riemann solver has a good capability of capturing shocks, which might be helpful to study climatology on exoplanets \citep{fromang2016shear}. The left-hand side of Equation~(\ref{equ:linear_sys}) is a tridiagonal linear system that can be solved via the simple Gaussian Elimination method. Thus, all terms in Equation~(\ref{equ:linear_sys}) are evaluable at the $n$-th time step. $\Delta \boldsymbol{Q}_i$ is solve by inverting Equation~(\ref{equ:linear_sys}) and finally the conservative variables at step $n+1$ are:
\begin{equation}
\label{equ:update}
    \boldsymbol{Q}_i^{n+1} = \Delta \boldsymbol{Q}_i + \boldsymbol{Q}_i^n.
\end{equation}

\subsection{VIC in the Multi-Stage Time Integration Scheme} \label{sec:rk3}

The previous section provided the formulation of the VIC scheme in a single-stage and forward-Euler time integration scheme. Here we discuss how to implement the VIC scheme in a multi-stage RK-type time integration scheme to achieve higher numerical stability and accuracy in time \citep{shu1988efficient}. The formulation of this implicit scheme is slightly different from the TVD method in the original paper \citep{li2019simulating}. Here, we list algorithms of third-order accurate Runge-Kutta time integration (RK3) method as an example. We also present both explicit and implicit algorithms for the comparison. The original explicit RK3 time integration scheme is:

\begin{equation}
\label{equ:exiplicit-RK3-stepI}
      \boldsymbol{Q}_{i}^{(1)} = \boldsymbol{Q}_{i}^{n} - \frac{\Delta t}{\Delta V_{i}}(\sigma_{i+1/2}\boldsymbol{F}_{i+1/2}^{n} - \sigma_{i-1/2}\boldsymbol{F}_{i-1/2}^{n}) + \Delta t (\boldsymbol{X}_{i}^{n}+\boldsymbol{Y}_{i}^{n}),
\end{equation}

\begin{equation}
\label{equ:exiplicit-RK3-stepII}
      \boldsymbol{Q}_{i}^{(2)} = \frac{3}{4}\boldsymbol{Q}_{i}^{n} + \frac{1}{4}\Big[ \boldsymbol{Q}_{i}^{(1)} - \frac{\Delta t}{\Delta V_{i}}(\sigma_{i+1/2}\boldsymbol{F}_{i+1/2}^{(1)} - \sigma_{i-1/2}\boldsymbol{F}_{i-1/2}^{(1)}) + \Delta t (\boldsymbol{X}_{i}^{(1)}+\boldsymbol{Y}_{i}^{(1)}) \Big],
\end{equation}

\begin{equation}
\label{equ:explicit-RK3-stepIII}
      \boldsymbol{Q}_{i}^{n+1} = \frac{1}{3}\boldsymbol{Q}_{i}^{n} + \frac{2}{3}\Big[ \boldsymbol{Q}_{i}^{(2)} - \frac{\Delta t}{\Delta V_{i}}(\sigma_{i+1/2}\boldsymbol{F}_{i+1/2}^{(2)} - \sigma_{i-1/2}\boldsymbol{F}_{i-1/2}^{(2)}) + \Delta t (\boldsymbol{X}_{i}^{(2)}+\boldsymbol{Y}_{i}^{(2)}) \Big],
\end{equation}
where the superscript (1) and (2) represent the first and the second intermediate states, respectively. We use the intermediate states of conservative variables, $\boldsymbol{Q}_{i}^{(1)}$ or $\boldsymbol{Q}_{i}^{(2)}$, to update the flux at the intermediate states, $\boldsymbol{F}_{i+1/2}^{(1)}$ and $\boldsymbol{F}_{i+1/2}^{(2)}$.

The implicit scheme replaces the explicit fluxes $\boldsymbol{F}$ and forcing $\boldsymbol{X}$ by their implicit counterparts. For example, the first step of the RK3 integration scheme becomes:

\begin{equation}
\label{equ:implicit-RK3-stepI}
      \frac{\Delta \boldsymbol{Q}_{i}^{n}}{\Delta t} + \frac{\sigma_{i+1/2}\boldsymbol{F}_{i+1/2}^{(1)} - \sigma_{i-1/2}\boldsymbol{F}_{i-1/2}^{(1)}}{\Delta V_{i}} = 
      \boldsymbol{X}_{i}^{(1)} + \boldsymbol{Y}_{i}^{n}.
\end{equation}
According to Equation~(\ref{equ:linear_sys}), $\Delta \boldsymbol{Q}_{i}^{n}$ is solved by inverting the tridiagonal system defined by Equation~(\ref{equ:implicit-RK3-stepI}). Then, the conservative variables are updated by the following equation,

\begin{equation}
\label{equ:implicit-RK3-stepI-add}
      \boldsymbol{Q}_{i}^{(1)} = \Delta \boldsymbol{Q}_{i}^{n} + \boldsymbol{Q}_{i}^{n}.
\end{equation}
Similarly, the other two intermediate stages are updated sequentially:

\begin{equation}
\label{equ:implicit-RK3-stepII}
    \begin{aligned}
      \frac{\Delta \boldsymbol{Q}_{i}^{(1)}}{\Delta t/4} + \frac{\sigma_{i+1/2}\boldsymbol{F}_{i+1/2}^{(2)} - \sigma_{i-1/2}\boldsymbol{F}_{i-1/2}^{(2)}}{\Delta V_{i}} =  \boldsymbol{X}_{i}^{(2)} + \boldsymbol{Y}_{i}^{(1)}, \\
      \boldsymbol{Q}_{i}^{(2)} =  \Delta \boldsymbol{Q}_{i}^{(1)} + \frac{1}{4} \boldsymbol{Q}_{i}^{(1)} + \frac{3}{4}\boldsymbol{Q}_{i}^{n},
    \end{aligned}
\end{equation}

\begin{equation}
\label{equ:implicit-RK3-stepIII}
    \begin{aligned}
      \frac{\Delta \boldsymbol{Q}_{i}^{(2)}}{2 \Delta t/3} + \frac{\sigma_{i+1/2}\boldsymbol{F}_{i+1/2}^{n+1} - \sigma_{i-1/2}\boldsymbol{F}_{i-1/2}^{n+1}}{\Delta V_{i}} = 
      \boldsymbol{X}_{i}^{n+1} + \boldsymbol{Y}_{i}^{(2)}, \\ 
      \boldsymbol{Q}_{i}^{n+1} = \Delta \boldsymbol{Q}_{i}^{(2)} + \frac{2}{3} \boldsymbol{Q}_{i}^{(2)} + \frac{1}{3}\boldsymbol{Q}_{i}^{n}.
    \end{aligned}
\end{equation}

\subsection{VIC in Three Dimensions} \label{sec:coupling}

In this section, we demonstrate how to combine the explicit scheme in the horizontal plane ($x$-$y$ plane) with the implicit scheme in the vertical $z$-direction. In particular, $\boldsymbol{H}$ is the flux in the vertical direction which is treated implicitly. The explicit scheme is computationally efficient in one time step but subject to the conditional numerical stability. The computational efficiency of an explicit non-hydrostatic atmosphere model is generally limited by the acoustic speed and atmospheric vertical resolution. On the contrary, the implicit scheme trades off the computational efficiency against numerical stability. Here we combine the advantages of the two numerical schemes to solve the 3D Euler equations.

Most of the existing atmosphere models with the vertically-implicit scheme use dimensionally-split method to compute horizontal and vertical terms separately \citep[e.g.,][]{ullrich2012operator,ullrich2012mcore,mendoncca2016thor}. Flux divergence in different directions is updated by different time integration schemes in some algorithms \cite[e.g.,][]{ullrich2012operator,bao2015horizontally}. Here, we use dimensionally-unsplit method to update conservative variables simultaneously in all $x$-, $y$-, and $z$-directions for each half time step and the final time step. The discretized equation using the dimensionally-unsplit formulae for 3D case is,

\begin{equation}
    \label{equ:3d-euler}
    \begin{aligned}
    \frac{\boldsymbol{Q}_{ijk}^{n+1} - \boldsymbol{Q}_{ijk}^{n}}{\Delta t} + &
    \frac{\sigma_{i+1/2;;}\boldsymbol{F}_{i+1/2;;}^{n} - \sigma_{i-1/2;;}\boldsymbol{F}_{i-1/2;;}^{n}}{\Delta V_{ijk}} + \frac{\sigma_{;j+1/2;}\boldsymbol{G}_{;j+1/2;}^{n} - \sigma_{;j-1/2;}\boldsymbol{G}_{;j-1/2;}^{n}}{\Delta V_{ijk}} + \\&
    \frac{\sigma_{;;k+1/2}\boldsymbol{H}_{;;k+1/2}^{n+1} - \sigma_{;;k-1/2}\boldsymbol{H}_{;;k-1/2}^{n+1}}{\Delta V_{ijk}} = 
    \boldsymbol{X}^{n+1}_{ijk} + \boldsymbol{Y}^{n}_{ijk},
    \end{aligned}
\end{equation}
where $\boldsymbol{F}$, $\boldsymbol{G}$, and $\boldsymbol{H}$ are numerical fluxes in $x$, $y$, and $z$ directions (or $\theta$, $\phi$, and $r$ in a spherical coordinate system), respectively. Note that only the vertical (or radial) flux gradient is treated implicitly. We simplified the notation of the subscript by using `$;$' to omit some cell-centered indices. For example, $\boldsymbol{Q}_{i+1/2;;}\equiv\boldsymbol{Q}_{i+1/2,j,k}$ 

Substituting Equations~(\ref{equ:flux-taylor}) and (\ref{equ:Roe-n+1}) into Equation~(\ref{equ:3d-euler}), the discretized Euler equations can be rewritten as,
\begin{equation}
\label{equ:3d-euler-HEVI}
\begin{aligned}
        \frac{1}{\Delta V_{ijk}}\Big[ c_{ijk}(\boldsymbol{Q}_{;;k-1}^{n+1} - 
        \boldsymbol{Q}_{;;k-1}^{n}) + (\frac{\boldsymbol{\mathcal{I}}}{\Delta t} - \boldsymbol{X'}_{ijk}^{n}  + a_{ijk})(\boldsymbol{Q}_{ijk}^{n+1} - \boldsymbol{Q}_{ijk}^{n}) + b_{ijk}(\boldsymbol{Q}_{;;k+1}^{n+1} - \boldsymbol{Q}_{;;k+1}^{n}) \Big] = 
        \boldsymbol{R}_{ijk} ,
\end{aligned}
\end{equation}
where $\boldsymbol{R}_{ijk}$ is,
\begin{equation}
\label{equ:Rijk}
\begin{aligned}
    \boldsymbol{R}_{ijk} = & -\frac{\sigma_{i+1/2;;}\boldsymbol{F}_{i+1/2;;}^{n} - \sigma_{i-1/2;;}\boldsymbol{F}_{i-1/2;;}^{n}}{\Delta V_{ijk}} -  \frac{\sigma_{;j+1/2;}\boldsymbol{G}_{;j+1/2;}^{n} -  \sigma_{;j-1/2;}\boldsymbol{G}_{;j-1/2;}^{n}}{\Delta V_{ijk}} \\&
    -\frac{\sigma_{;;k+1/2}\boldsymbol{H}_{;;k+1/2}^{n} - \sigma_{;;k-1/2}\boldsymbol{H}_{;;k-1/2}^{n}}{\Delta V_{ijk}} 
    + \boldsymbol{X}^{n}_{ijk} +  \boldsymbol{Y}^{n}_{ijk} \\
    = & \frac{\boldsymbol{Q}_{ijk}^{(e)}-\boldsymbol{Q}_{ijk}^{n}}{\Delta t}.
\end{aligned}
\end{equation}

Denoted by $\boldsymbol{Q}_{ijk}^{(e)}$ is the conservative variables predicted by the explicit formulation. The time integration scheme for the 3D case is similar to the 1D time integration scheme in Section~\ref{sec:rk3}. We observe that, for some 3D simulations, the horizontal CFL number can reach as large as $1.6$ (i.e., $c_{s} \Delta t/\mathrm{min}(\Delta x, \Delta y) \approx 1.6$) with the implicit RK3 time integration scheme. The reason for this ``ultra-stable'' behavior is probably due to the unsplit nature of our implicit scheme such that relaxing the theoretical CFL limit in one direction helps to improve numerical stability in the other directions.
A rigorous numerical study of the VIC scheme will be devoted to forthcoming studies. Finally, we summarize the whole implicit scheme in the following step:

\begin{enumerate}
    \item Perform an explicit forward step and calculate $\boldsymbol{R}_{ijk}$ according to equation~(\ref{equ:Rijk}).
    \item Calculate coefficients of the block tridiagonal matrix ($a_{ijk}$, $b_{ijk}$, $c_{ijk}$) according to equations~(\ref{equ:a}), (\ref{equ:b}), and (\ref{equ:c}).
    \item Solve the block tridiagonal matrix column by column defined by equation~(\ref{equ:3d-euler-HEVI}).
    \item Update the conserved variables according to equation~(\ref{equ:update}).
    \item Repeat the preceding steps for each stage of a multi-stage time integration scheme similar to equations~(\ref{equ:implicit-RK3-stepI-add}), (\ref{equ:implicit-RK3-stepII}) and (\ref{equ:implicit-RK3-stepIII})
\end{enumerate}

Because steps 1, 4, and 5 are needed for any explicit integration scheme, an existing explicit model can employ our implicit scheme by simply adding additional implicit correction steps outlined in step 2 and step 3, which makes the scheme extremely flexible and versatile.

\section{Conservation Laws} \label{sec:conservation-law}

Conservation laws are important for atmospheric simulations. Built on top of the FVM framework, whose explicit scheme has already been well-designed for conservation laws, our VIC scheme can rigorously satisfy the conservation of total mass and energy and perhaps total momentum with appropriate boundary conditions and coordinate systems. The conservation of angular momentum under spherical coordinates cannot be numerically guaranteed since we solve momentum equations instead of angular momentum equations. In this section, we particularly discuss conservation laws under a Cartesian framework while disregarding the body force (i.e., gravitational acceleration, etc.).

The conservation of total mass, momentum, and energy can be guaranteed by the intrinsic relationship between the elements in the matrix of the linear system, $a_{i}$, $b_{i}$, and $c_{i}$, which satisfy $b_{i-1} + a_{i} + c_{i+1} = 0$. The total mass, momentum, and energy changes from the current time step to the next time step can be mathematically written as,

\begin{equation}
\label{equ:conservation}
    \sum_{i=1}^m \boldsymbol{Q}_{i}^{n+1} - \sum_{i=1}^m \boldsymbol{Q}_{i}^{n} =  \sum_{i=1}^m (\boldsymbol{Q}_{i}^{n+1} - \boldsymbol{Q}_{i}^{n}) = \sum_{i=1}^m \Delta \boldsymbol{Q}_i = 0,
\end{equation}
where $m$ is the total number of cells in the vertical direction. $i = 1$ and $i = m$ are the first and the last cell in the domain, respectively. $i = 0$ and $i = m+1$ are used to denote the location of ghost cells. Ghost cells are out of the domain and contain information about boundary conditions. Therefore, $\sum_{i=1}^m \Delta \boldsymbol{Q}_{i}$ can be acquired from the linear system, Equation~(\ref{equ:linear_sys}), by adding up all equations,

\begin{equation}
\label{equ:compressed-linear}
    \frac{1}{\Delta t} \sum_{i=1}^{m} \Delta \boldsymbol{Q}_i = 
    -B_{0}(\Delta \boldsymbol{Q}_0) - B_{m}(\Delta \boldsymbol{Q}_m) - \sum_{i=2}^{m-1} \big[ (b_{i-1} + a_{i} + c_{i+1}) \Delta \boldsymbol{Q}_i \big] + \frac{(\boldsymbol{F}_{m+1/2}^{n} - \boldsymbol{F}_{1/2}^{n})}{\Delta x},
\end{equation}
where $B_{0}(\Delta \boldsymbol{Q}_0)$, $B_{m}(\Delta \boldsymbol{Q}_m)$, $\boldsymbol{F}_{1/2}^{n}$, and $\boldsymbol{F}_{m+1/2}^{n}$ are determined by boundary conditions. As previously mentioned, with $b_{i-1} + a_{i} + c_{i+1} = 0$ for $i = 2, 3, 4, ..., m-1$, the third term on the right-hand side of the equation is zero. Then, as previously presented, conservation laws are purely decided by boundary conditions. 

Physically, only the double-periodic boundary condition can guarantee the conservation of total mass, momentum, and energy in the domain. The reflecting boundary condition only conserves total mass and energy. Our VIC scheme can be applied to a vairety of boundary conditions by modifying the linear system in Equation~\ref{equ:linear_sys}. Similar to the other vertically implicit schemes \citep{mendoncca2016thor}, linear equations about the first cell (i.e., the first row of the linear system) and the last cell (i.e., the last row of the linear system) in the domain interior are modified to satisfy different boundary conditions.

In Appendices~\ref{sec:reflecting-prove} and ~\ref{sec:double-periodic-prove}, we present the detailed proof of mass and energy conservation under the reflecting boundary condition and mass, momentum, and energy conservation under the double-periodic boundary condition, respectively. The reflecting boundary condition (i.e., no-flux or free-slip boundary condition) is the most commonly applied to both terrestrial and gaseous planetary-atmosphere simulations.

\section{Benchmark Test Validation} \label{sec:benchmark}

In this section, we present the code performance and verification tests of the VIC scheme against several standard numerical benchmark tests. Benchmark simulations can test if the VIC scheme can remain stable under large time steps (i.e., relaxed CFL number), track turbulent motions, resolve meteorologically significant gravity waves, and produce large-scale dynamics. They can also validate the numerical performance of conservation laws. We provide the simulation results of fully-explicit numerical schemes as a comparison to the results using the VIC scheme. We present the simulation results of the linearized acoustic wave test, Straka sinking bubble test \citep{straka1993numerical}, Robert rising bubble test \citep{robert1993bubble}, gravity wave test \citep{skamarock1994efficiency}, Held-Suarez atmospheric experiment \citep{held1994proposal}, and shallow hot Jupiter test. These tests are summarized in Table~\ref{table:1}.

\begin{table}[t]
\begin{threeparttable}
\caption{Local and global benchmark cases in this study.} \label{table:1}
\centering
\begin{tabular}{ p{0.2\linewidth}p{0.15\linewidth}p{0.35\linewidth}p{0.2\linewidth} }
\hline
\multicolumn{1}{c}{Test Name} & \multicolumn{1}{c}{Dimensions} & \multicolumn{1}{c}{Test Purpose} & \multicolumn{1}{c}{Boundary Conditions} \\
\hline
Linearized acoustic wave          & \multicolumn{1}{c}{1D local} & Testing spurious numerical noise damping    & \multicolumn{1}{c}{II}      \\ 
Straka sinking bubble             & \multicolumn{1}{c}{2D local} & Resolving nonlinear density current   & \multicolumn{1}{c}{I}     \\ 
Robert rising bubble              & \multicolumn{1}{c}{2D local} & Tracking weak turbulence                    & \multicolumn{1}{c}{I}        \\ 
Localized gravity wave            & \multicolumn{1}{c}{2D local} & Resolving gravity waves                     & \multicolumn{1}{c}{I,II}     \\ 
Held-Suarez                       & \multicolumn{1}{c}{3D global} & Generating thermal wind                    & \multicolumn{1}{c}{I,II,III} \\
Shallow hot Jupiter               & \multicolumn{1}{c}{3D global} & Generating super-rotating jet              & \multicolumn{1}{c}{I,II,III} \\ 
\hline
\end{tabular}
\begin{tablenotes}
     \item \textbf{Notes}: Three types of boundary conditions in $\rm Athena^{++}$ are used. I: reflecting; II: double-periodic; III: polar wedge. See \citet{stone2020athena++} for the scheme description. 
    \end{tablenotes}
\end{threeparttable} 
\end{table}

The simulation results of the Straka sinking bubble test and Robert rising bubble test in the original papers only show results with the aspect ratio equals to one (i.e., $\Delta x/\Delta z = 1$) \citep{straka1993numerical,robert1993bubble}, which does not usually occur in a more realistic atmospheric simulation (i.e., $\Delta x/\Delta z \gg 1$). In this case, we cannot test the implicit scheme with a very large time step because of the time step limitation in the horizontal direction. Here, we will present the simulation results with the aspect ratio of 10 in this paper to validate the performance of the code in a large time step. The CFL numbers are computed differently in the explicit and VIC schemes. The time step is computed by $\Delta t_{exp} = CFL\cdot \mathrm{min}(\Delta z,\Delta y,\Delta z)/c_{s}$ in the explicit scheme, on the contrary, the time step is computed by $\Delta t_{VIC} = CFL\cdot \mathrm{min}(\Delta x, \Delta y)/c_{s}$ in the VIC scheme.

\subsection{Linear Wave Test: Acoustic Wave Damping} \label{sec:linear-wave}

The VIC scheme relaxes the CFL limitation by damping fast acoustic waves. In most benchmark tests that are designed for atmospheric simulations, the vertical wind velocity is slower than the acoustic speed by more than two orders of magnitude, meaning that the time step is predominantly limited by the acoustic wave speed. The VIC scheme resolves this issue by imposing numerical diffusivity which suppresses the accumulation of the spurious numerical noise.

We validate the model by simulating the propagation of the linearized acoustic wave. This test is very similar to the linear wave convergence test in \cite{stone2008athena} but we focus on how numerical diffusivity damps the amplitude of the acoustic wave. Linear acoustic waves are launched by an initial pressure perturbation. The polytropic index, $\gamma$, is set as 5/3. Density, pressure, and velocity of the uniform background are initialized with 1.0 Pa, 0.6 $\rm g \; cm^{-3}$, and 0 $\rm m \; s^{-1}$, respectively. The corresponding acoustic speed, $c_{s}$, is 1 $\rm m \; s^{-1}$. The wave amplitude is set as $\rm 10^{-3}$ Pa. The wave propagates in a 1D double-periodic tube whose wavelength equals to the length of the domain. We simulate the propagation of acoustic waves with different time steps (i.e., CFL numbers) and collect the density profiles after 1 second in the simulation.

\begin{figure}
    \centering
    \includegraphics[width=1.0\textwidth]{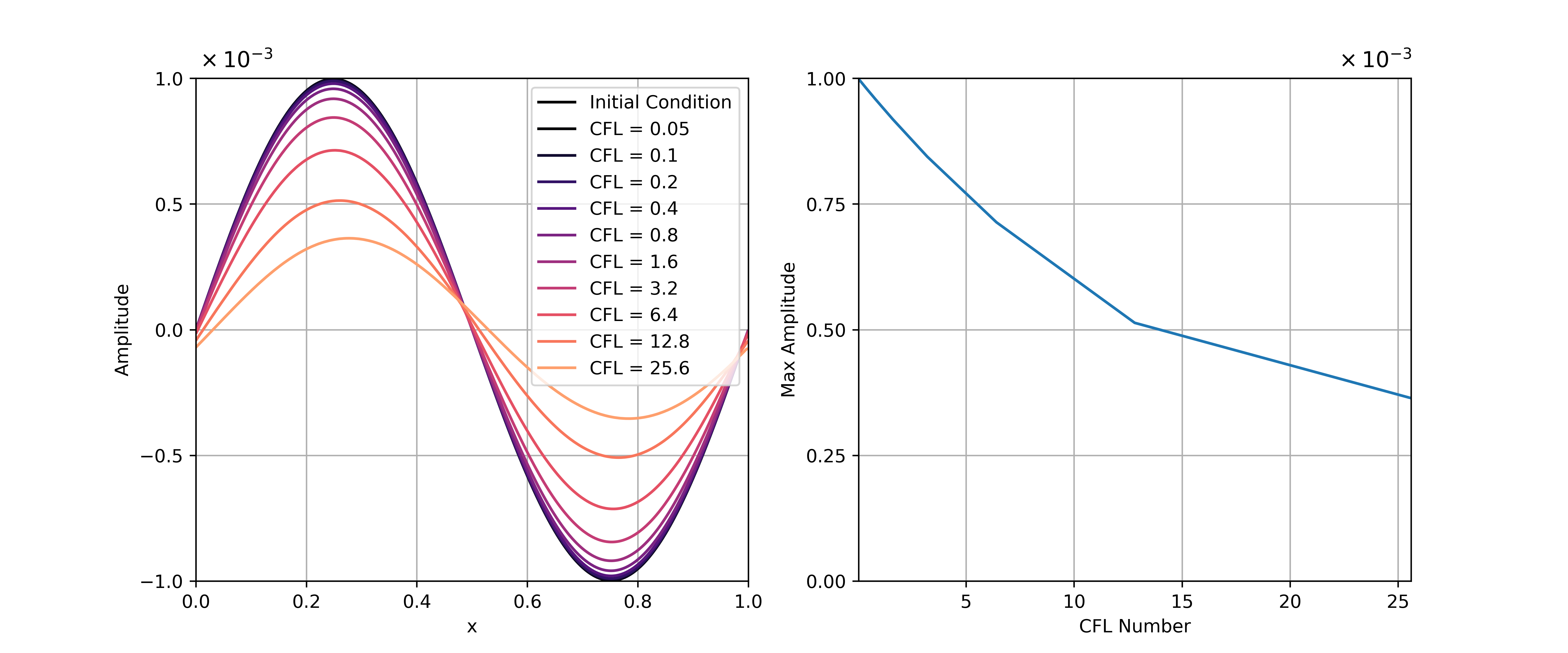}
    \caption{The left plot shows the wave amplitude damping and dispersion errors with different CFL numbers from 0.05 to 25.6 in a 1D tube with 512 grids. The right plot shows the wave amplitude damping as a function of the CFL number.}
    \label{fig:damp}
\end{figure}

The left plot in Figure~\ref{fig:damp} shows the density profiles after one wave period with different CFL numbers but the same resolution. The numerical diffusivity damps the acoustic wave amplitude significantly for large CFL numbers. The large CFL number simulations also cause the dispersion error for the acoustic waves. The right plot in Figure~\ref{fig:damp} shows that the damping rate is almost growing linearly with increasing the CFL number. We can infer the increase of the damping rate by analyzing Equation~(\ref{equ:linear_sys}). The numerical diffusivity is imposed by the off-diagonal elements in Equation~(\ref{equ:linear_sys}). They are independent of the changing time step but only dependent on the spatial resolution. When we fix the spatial resolution and increase $\Delta t$, the significance of the off-diagonal elements becoming more and more important because $1/\Delta t$ in diagonal terms becomes smaller and smaller. Therefore, the damping rate increases linearly with increasing CFL number.

\subsection{Straka Sinking Bubble Test} \label{sec:straka}

\begin{figure}[b]
    \centering
    \includegraphics[width=1.0\textwidth]{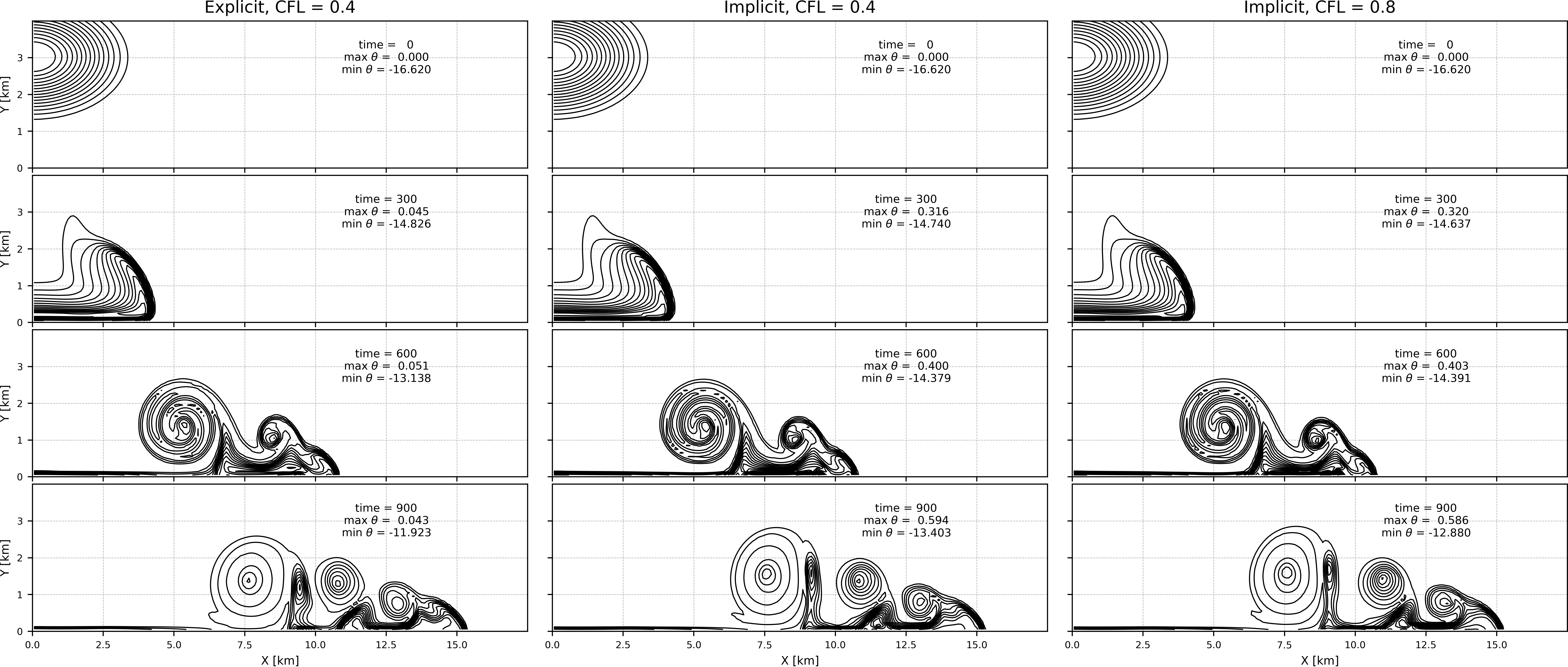}
    \caption{Nearly inviscid simulation results of the Straka sinking bubble test with different numerical schemes and CFL numbers. The left column shows the result of the explicit SNAP model with CFL $\rm = 0.4$; the middle and right columns show the results of the SNAP model with the VIC scheme. CFL numbers of the numerical solutions in the middle and the right columns are 0.4 and 0.8, respectively. The plotted domain size is $[0, 17.5] \times [0, 4]$ in kilometers. The spatial resolutions of the three cases are the same, which are $ \Delta x = 100$ m and $\Delta z = 100$ m.}
    \label{fig:straka-ratio1}
\end{figure} 

We have shown that the numerical diffusivity imposed by the VIC can damp the acoustic wave amplitude from the linearized acoustic wave test. This helps the VIC to achieve numerical stability when a large time step is used. However, large numerical diffusivity may hinder the ability to resolve turbulence. It is important to validate the VIC scheme's ability of correctly tracking turbulent motions. In this and the next section, we will focus on the VIC scheme's capability of resolving non-linear density currents and turbulent motions.

The Straka sinking bubble test is a standard benchmark test, which is designed for the validation of non-hydrostatic atmospheric dynamical cores \citep{straka1993numerical}. This case simulates the fluid motion in a nonlinear density current generated by a sinking cold bubble. Several physical processes (i.e., Kelvin-Helmholtz instability) should be produced after the bubble dropped on the surface \citep{straka1993numerical}. The simulation is carried out in a 25.6 km by 6.4 km closed domain. The initial background temperature structure is set as dry adiabat in the vertical direction. The reference surface pressure is set as 1 bar ($\rm 10^5$ Pa). A cold bubble, whose center is 15 K colder than the local background temperature, is put aloft. The cold bubble is expected to drop to the surface and generate several non-linear density currents. In the original test, the explicit diffusion is also applied to momentum and energy equations to guarantee convergence \citep{straka1993numerical}. In this work, however, we do not apply the diffusion terms in our equations in order to check the performance difference between our VIC scheme and the explicit scheme. The initial temperature is given by:

\begin{equation}
    \Delta T = 
    \left \{
    \begin{tabular}{cc}
         &  $ 0 $, if $L > 1$, \\
         &  $ -15[\cos{(\pi L)} + 1]/2 $, if $L \le 1$,
    \end{tabular} 
    \right.
\end{equation}
where $L = \{ [(x-x_c)/x_r]^2 + [(z-z_c)/z_r]^2 \}^{1/2}$, $ x_c = 0$ km, $ x_r = 4$ km, $ z_c = 3$ km, and $ z_r = 2$ km.

For the explicit test, we adopt the same numerical techniques as adopted in \cite{li2019simulating} using the Low Mach Number Approximate Riemann Solver (LMARS), WENO5 reconstruction scheme, and RK3 time integration scheme, to ensure our result is comparable with the numerical solutions in \cite{li2019simulating}. We first investigate our VIC scheme's capability of capturing the position of the bubble debris and Kelvin-Helmholtz rotors. Then, we compare the potential temperature difference between the results of the VIC scheme and the explicit SNAP from \cite{li2019simulating} to estimate the role of the numerical diffusivity. 

\begin{figure}
    \centering
    \includegraphics[width=1.0\textwidth]{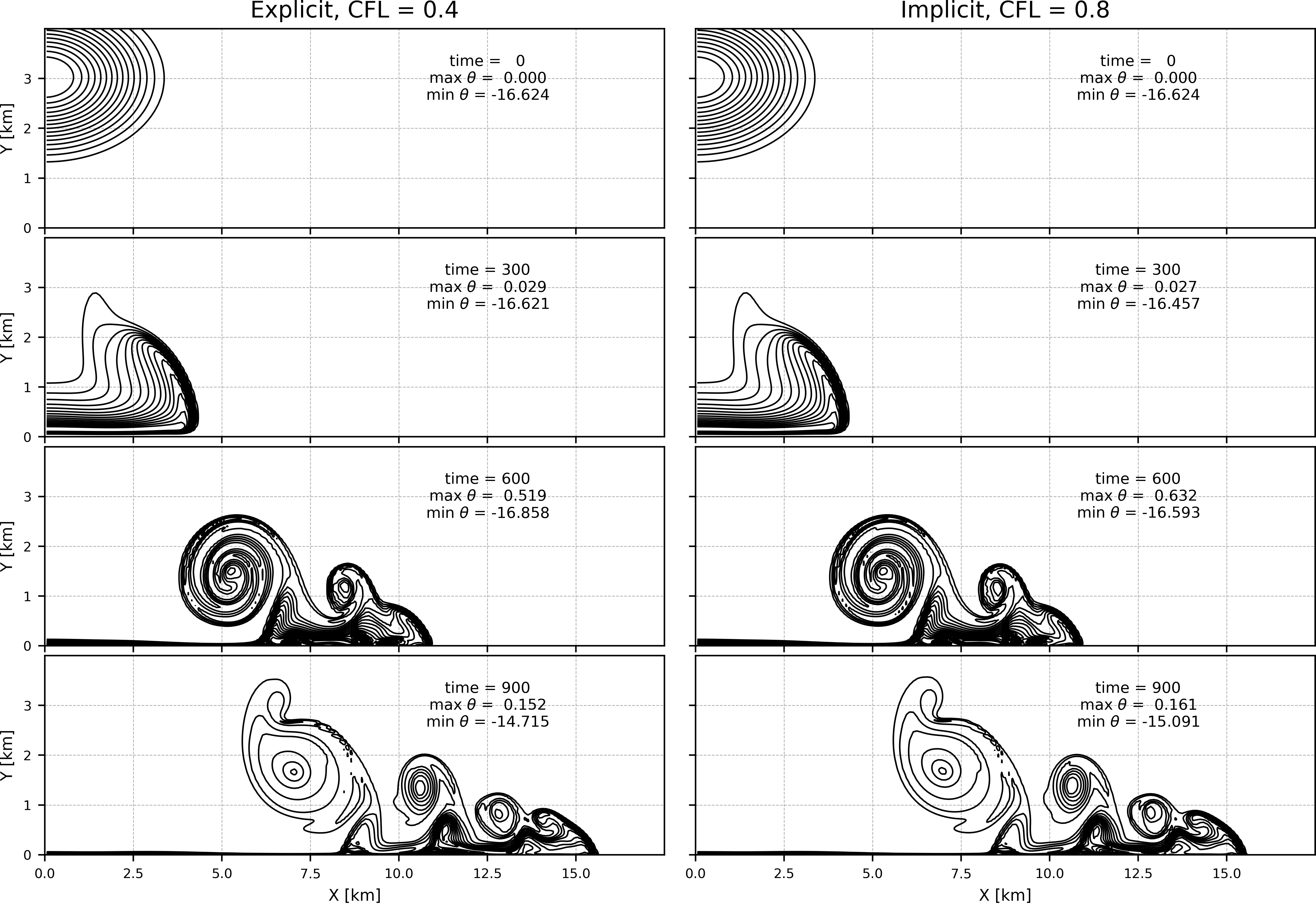}
    \caption{Another set of nearly inviscid simulation results of the Straka sinking bubble test with the aspect ratio of 10. The left column shows the result of the explicit SNAP scheme with CFL $\rm = 0.4$; the right column shows the numerical solution using the VIC scheme with CFL $\rm = 0.8$. The spatial resolutions of these two cases are the same, which are $\Delta x = 100$ m and $\Delta z = 10$ m.}
    \label{fig:straka-ratio10}
\end{figure}

Simulation results at $t = 900$ s with an aspect ratio of 1 (i.e., $\Delta x/\Delta z = 1$) are presented in Figure~\ref{fig:straka-ratio1}. The maximum CFL numbers are 0.5 and 1.0 for the explicit and implicit schemes, respectively. Here, we present and compare the results of using the explicit scheme with CFL $\rm = 0.4$, the VIC scheme with CFL $\rm = 0.4$, and the VIC scheme with CFL $\rm = 0.8$ in Figure~\ref{fig:straka-ratio1}. Simulation results show that they generally converge to the same solution morphologically. Three Kelvin-Helmholtz rotors are correctly produced in all three cases. The position of the density currents' outer limb is about 15300 m in all results. A slight potential temperature difference can be seen from the plots. They are caused by numerical diffusion. The potential temperature deviations of the coldest part of the bubble debris are roughly 11.9 K, 13.4 K, and 12.9 K for three cases, respectively. For the explicit cases, the numerical diffusivity becomes smaller when the CFL number becomes larger; whereas the numerical diffusivity caused by the VIC scheme becomes larger when the time step becomes larger. This is expected because the non-diagonal terms, which performs as the numerical diffusivity, are more dominant than the diagonal terms when time step becomes larger.

\begin{figure}
    \centering
    \includegraphics[width=1.0\textwidth]{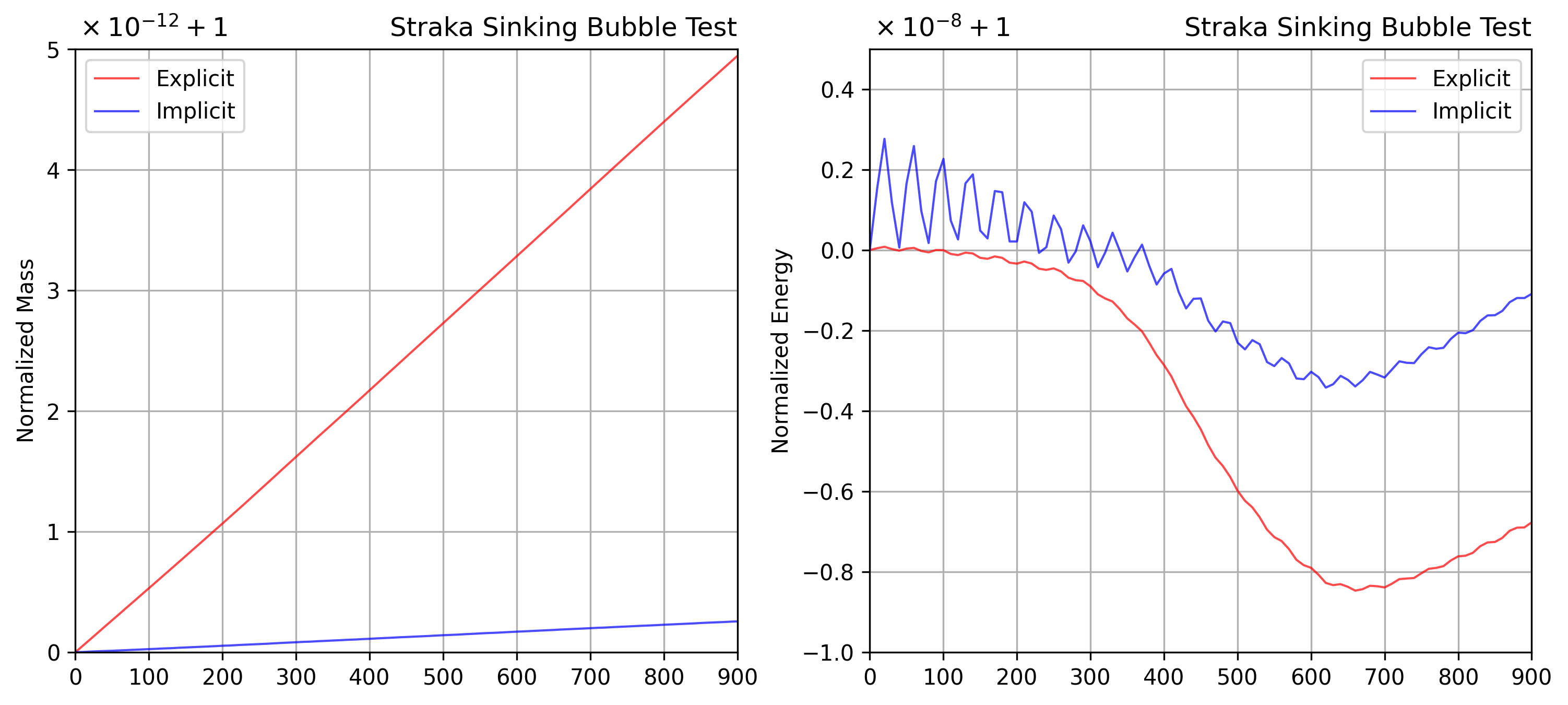}
    \caption{Temporal evolution of total mass (left) and energy (right) normalized by the initial values from explicit (red) and implicit (blue) simulations in Figure~\ref{fig:straka-ratio10}. The total energy calculation includes internal energy, kinetic energy, and gravitational potential. Explicit and implicit results are labeled with red and blue colors, respectively.}
    \label{fig:strak-conservation}
\end{figure}

We also present two innovative simulation results with the large aspect ratio (i.e., $\Delta x/\Delta z = 10$) in Figure~\ref{fig:straka-ratio10}. The large aspect ratio of the vertical resolution and horizontal resolution allows us to use a much larger CFL number for the VIC scheme. The VIC scheme adopts a time step which is larger than the time step in the explicit case by a factor of 20. The dynamical evolution of the bubble is not affected by the usage of a much larger time step in the VIC scheme but the computational efficiency is significantly improved. It takes about 31 minutes for the explicit scheme to finish the computation on Pleiades with 32 CPU cores (Sandy Bridge processors with frequency is 2.6 GHz). The VIC scheme, on the other hand, only takes about 3 minutes, illustrating that the computational efficiency is improved by one order of magnitude without influence on dynamics.

Figure~\ref{fig:strak-conservation} shows the evolution of total mass and energy as a function of time for cases using the large aspect ratio. This result is presented to show the performance of conservation law on both the VIC and the explicit scheme. It shows that total mass linearly increases as a function of time on the machine precision level in both explicit and implicit cases. The simulation using the VIC scheme has a smaller mass error compared with the explicit one due to the less workload and fewer time steps. The fractional variation of total energy in both cases is about $\rm 10^{-9}$ of the initial value. 

In sum, the Straka sinking bubble test shows that, despite the small and dynamically trivial potential temperature differences among these cases, our VIC scheme can reproduce numerical results in the explicit integration. It also shows that the VIC scheme can well conserve total mass and energy in a closed system (e.g., with reflecting boundary conditions). 

\subsection{Robert Rising Bubble Test} \label{sec:robert}

\begin{figure}
    \centering
    \includegraphics[width=1.0\textwidth]{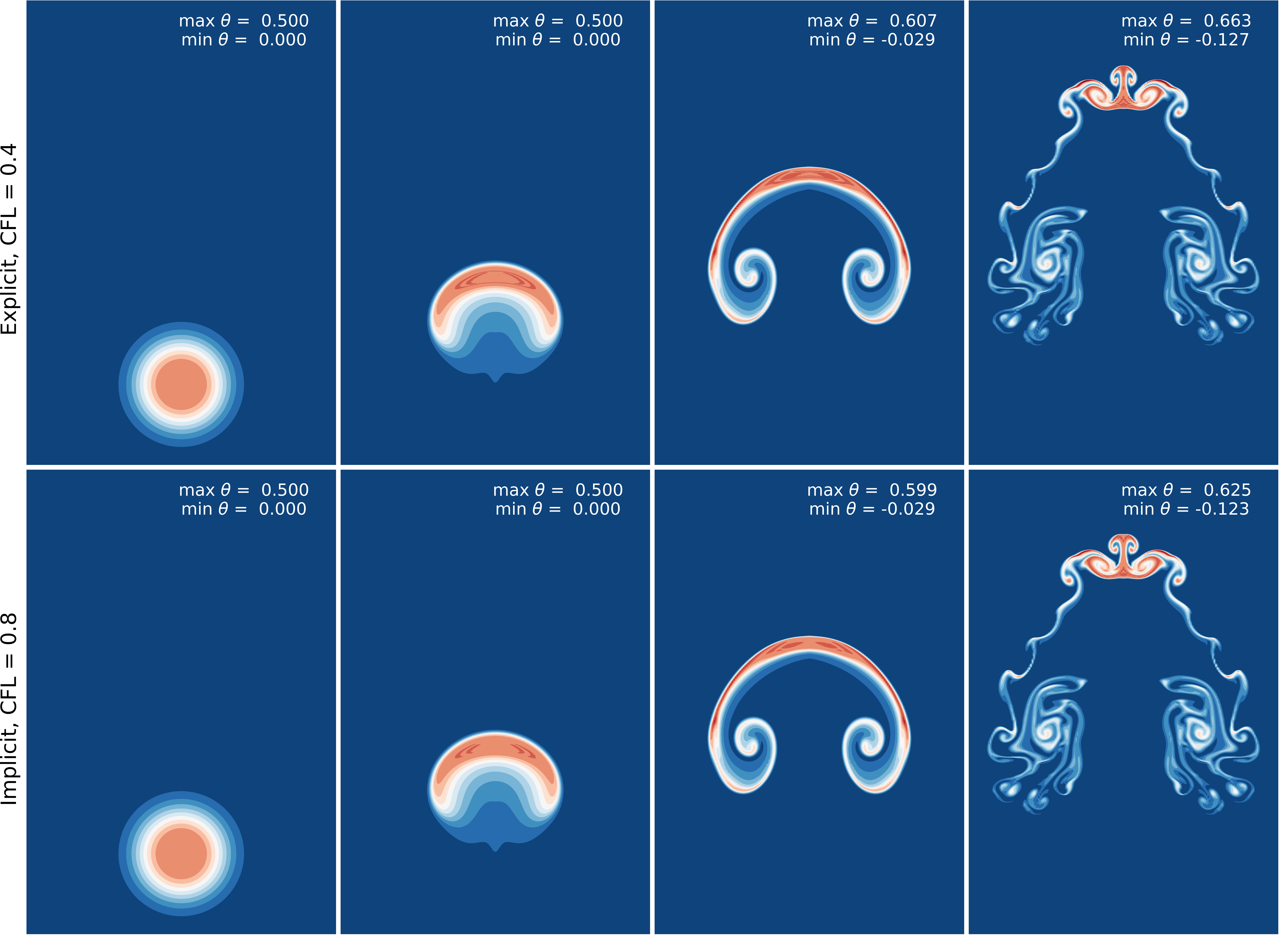}
    \caption{Simulation results of the Robert rising bubble test with the aspect ratio of 10. The horizontal resolution is 5 m and the vertical resolution is 0.5 m. Top panels show the explicit simulation results with the CFL number of 0.4 (i.e., $\Delta t \sim 5\times10^{-4}$ s). Bottom panels show the implicit simulation results with the CFL number of 0.8 (i.e., $\Delta t \sim 1\times10^{-2}$ s). The CFL number is computed from the horizontal resolution in the VIC scheme.}
    \label{fig:robert-aspect-10}
\end{figure}

The third benchmark test is the Robert rising bubble test, which is designed to test the model performance under a weak forcing. The simulation results of the Straka sinking bubble test show that the VIC scheme can correctly track the fluid motion in the gravity-dominant regime. However, further validation of turbulence tracking is necessary for the regime with the weak buoyant forcing. The Robert rising bubble test is a suitable benchmark test for our needs \citep{robert1993bubble}. This benchmark test has some practical implications because there are ubiquitous convective air parcels in the moist convective layer (i.e., troposphere) and, unlike the situation in the Straka bubble test, atmospheric convection is usually triggered by a small temperature perturbation (i.e., less than 1 K). The simulation is initialized with a Gaussian-shaped warm bubble and finished with a very turbulent snapshot \citep{robert1993bubble}.

Similar to the Straka benchmark test, the ambient atmosphere temperature is adiabatic with 303.15 K at the surface in a 1.5 km by 1 km closed box. The central temperature of the Gaussian-shaped bubble is set to be 0.5 K warmer than the background. The analytical formulation of the initial temperature profile is given by

\begin{equation}
    \Delta \theta = 
    \left \{
    \begin{tabular}{lc}
         &  $ A $ , if $r \leq a$,\\
         &  $ A\,e^{{-(r-a)^2}/{s^2}}$, if $r > a, $
    \end{tabular} 
    \right.
\end{equation}
where $r^2 = (x-x_{0})^2 + (z-z_{0})^2$, $A = 0.5$ K, $a = 50$ m, $s = 100$ m, $x_0 = 500$ m, and $z_0 = 260$ m. The warm bubble is expected to rise upward due to buoyancy. \cite{robert1993bubble} shows that the uplifting motion stops at about 18 minutes (simulation time) and the Kelvin-Helmholtz instability is developed at the tail of the warm air parcel \citep{robert1993bubble}. A fine resolution is necessary to be applied, at least 5 m in horizontal or vertical, to resolve the Kelvin-Helmholtz instability at 18 minutes \citep{robert1993bubble}.

We present the simulation results with a large aspect ratio (i.e., $\Delta x/\Delta z = 10$) in Figure~\ref{fig:robert-aspect-10}. Similar to the Straka sinking bubble test, this set-up allows us to test the performance of our VIC scheme in a time step of an order of magnitude larger. The implicit numerical solutions are mostly identical to explicit results. Our VIC scheme can reproduce the same turbulent patterns and Kelvin-Helmholtz instability as in the explicit solution at 18 minutes. The predominant difference is still the potential temperature. The head of the bubble showing in the implicit result is also slightly larger than the explicit one. In general, this simulation result agrees with the solutions reported in the previous studies \citep{robert1993bubble,chen2013control,guerra2016high,li2019simulating}. Although our implicit scheme possesses a strong numerical diffusion for large CFL numbers, the numerical diffusivity specifically suppresses the growing numerical noise of acoustic wave without any significant influence on turbulent flows induced by the very weak forcing.

The simulation efficiency is significantly improved by the VIC scheme for large-aspect-ratio simulations. Both explicit and implicit simulations use 50 CPU cores on Pleiades (Sandy Bridge processors) for the computation. The VIC scheme allows a larger time step by a factor of 20 compared to the explicit case. The implicit scheme is faster than the explicit scheme by a factor of eight in this case. 

\subsection{Gravity Wave Test}

\begin{figure}
    \centering
    \includegraphics[width=0.8\textwidth]{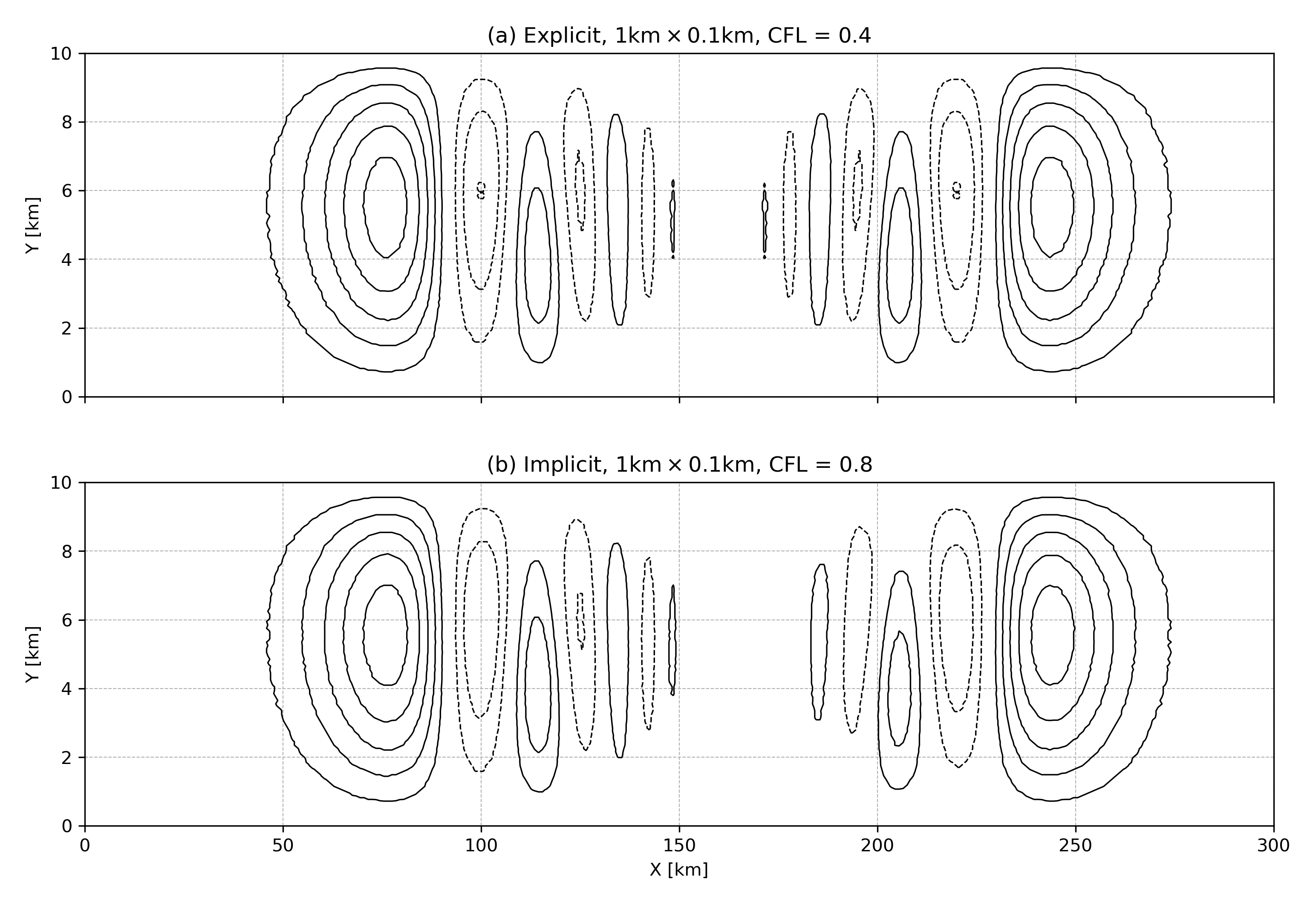}
    \caption{Potential temperature contours for the gravity wave test at $t$ = 3000 s. The solid contours refer to positive potential temperature deviation from the background, while the dashed contours represent negative values. Contours are plotted with an interval of $5\times 10^{-4}$ K. The  numerical schemes, CFL numbers, and spatial resolutions are shown in the title of each plot.}
    \label{fig:sk-gravity-wave}
\end{figure}

It is important for an atmospheric dynamical core to correctly resolve the dispersion relation of gravity waves because it preserves the information on how much energy and momentum are conveyed by waves. For example, inertia gravity waves play important roles in the momentum budget in the middle atmosphere, such as generating the Quasi-Biennial Oscillation (QBO) in Earth stratosphere \citep{baldwin2001quasi} and the Quasi-Quadrennial Oscillation (QQO) in Jupiter's stratosphere \citep{cosentino2017new}. Here, we present simulation results of a gravity wave benchmark test, which were first introduced by \cite{skamarock1994efficiency}. 

The model setup is similar to \cite{skamarock1994efficiency} and \cite{chen2013control}. We adopt a stable and hydrostatic atmosphere in a 300 km $\times$ 10 km domain with a small initial potential temperature perturbation to launch a train of gravity waves. The background atmosphere is initialized with a constant buoyancy frequency (i.e., Brunt-V$\ddot{\mathrm{a}}$is$\ddot{\mathrm{a}}$l$\ddot{\mathrm{a}}$ frequency), $N = 10^{-2}$ $\rm s^{-1}$. The surface pressure is set as 1 bar. The analytical potential temperature perturbation is

\begin{equation}
\Delta\theta = \Delta\theta_{0} \frac{\sin(\pi z/H)}{1 + (x-x_{c})^{2}/a^{2}},  
\end{equation}
where $\Delta \theta_{0} = 0.01$ K is the maximum potential temperature perturbation; $x_{c} = 100$ km, $a = 5$ km, and $H$ is the domain height, 10 km. A prescribed horizontal wind profile, $u = 20$ $\rm m \; s^{-1}$, is initialized as the background wind for the wave propagation. Thus one set of gravity waves propagates eastward with respect to the reference wind field, while another set of waves is quasi-stationary.

The reflecting boundary condition is adopted for the lower and upper boundaries. The double-periodic condition is used in the horizontal direction. Our set up is different from the original test \citep{skamarock1994efficiency} but is close to the simulations in \cite{chen2013control} and \cite{bao2015horizontally}. We simulate non-hydrostatic wave activities with density variations instead of using an anelastic assumption in the original paper \citep{skamarock1994efficiency}. Furthermore, similar to \cite{chen2013control}, we do not include Coriolis forces in the simulation to simplify the problem.

The simulation results in a train of dispersive gravity waves at 3000 s as shown in Figure~\ref{fig:sk-gravity-wave}. Both explicit and implicit schemes can accurately resolve the phase and dispersion relation of the wave. This is crucial for simulating the propagation of momentum and energy. On the other hand, gravity waves with a small amplitude near $\rm \sim$160 km are not resolved using our VIC scheme, indicating that the VIC scheme has a difficulty resolving very weak perturbations (i.e., $\Delta \theta \sim 0.001$ K). Furthermore, the explicit scheme can preserve the symmetrical structure of waves, but it seems that the performance of our VIC scheme is not as good as the explicit scheme under weak perturbations (i.e., at about $\rm \sim$ 170 km in Figure~\ref{fig:sk-gravity-wave}b).

\subsection{Held-Suarez Benchmark for Earth Atmosphere} \label{sec:hs94}

\begin{figure}
    \centering
    \includegraphics[width=0.8\textwidth]{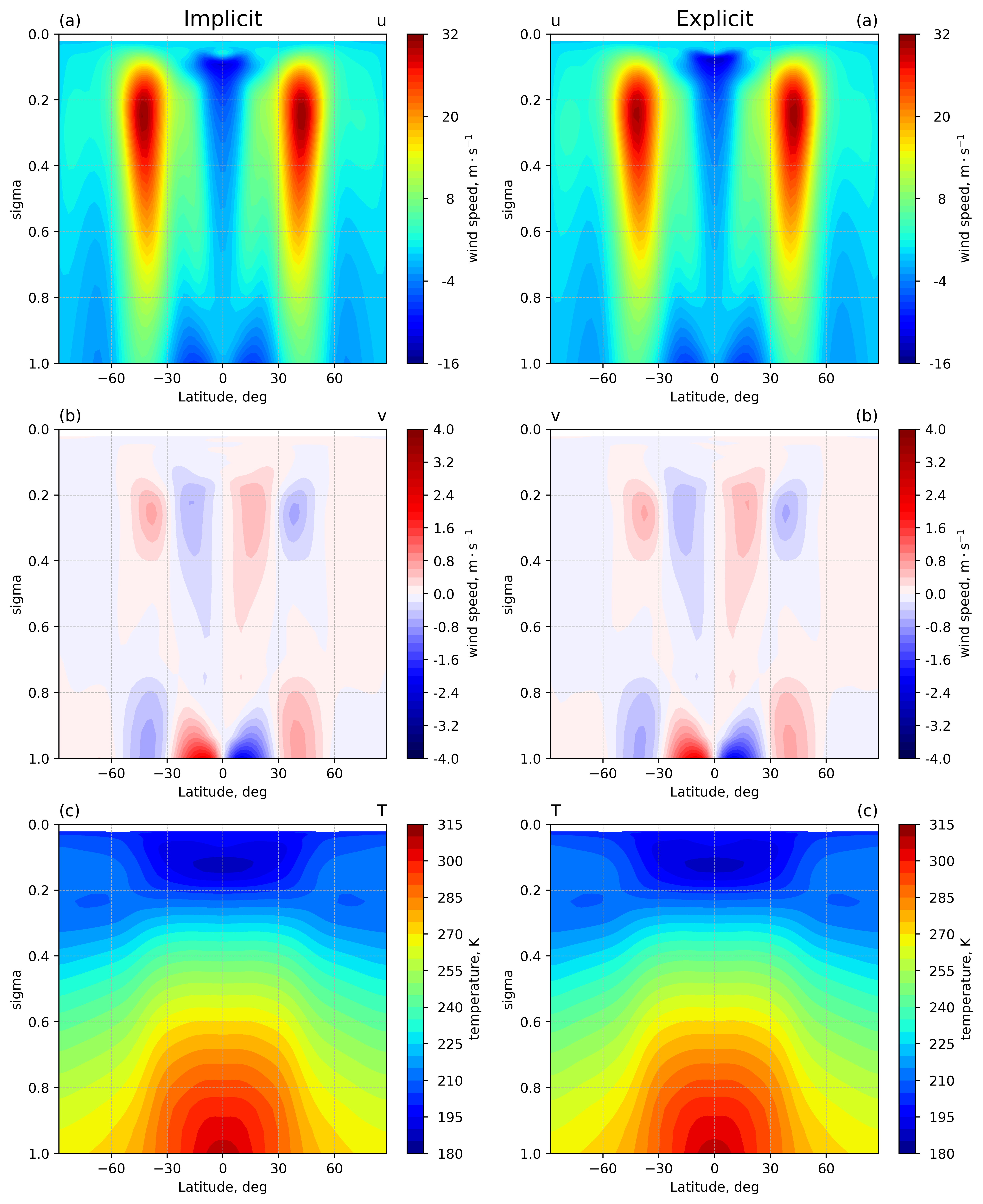}
    \caption{Explicit and implicit simulation results of the Held-Suarez benchmark averaged over 1000 Earth days. Zonal-mean zonal wind, meridional wind, and temperature are shown in the (a), (b), and (c) rows, respectively.}
    \label{fig:HS94-1}
\end{figure}

\begin{figure}
    \centering
    \includegraphics[width=0.8\textwidth]{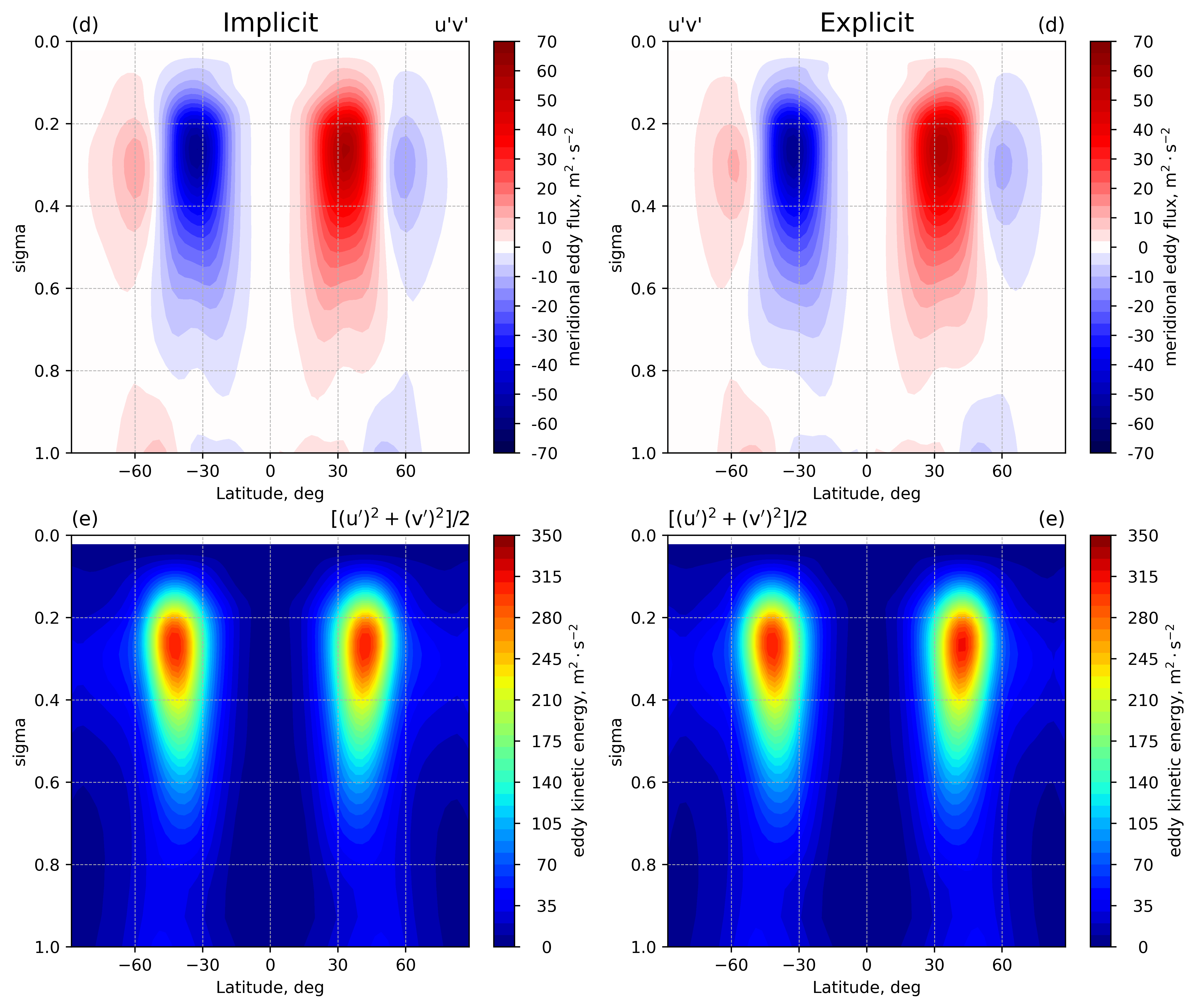}
    \caption{Explicit and implicit simulation results of the Held-Suarez benchmark averaged over 1000 Earth days. Zonal-mean meridional eddy momentum flux and eddy kinetic energy are displayed in (d) and (e), respectively.}
    \label{fig:HS94-2}
\end{figure}

The Held-Suarez experiment is designed for the intercomparison of GCM dynamical cores to produce the global-scale atmospheric features of an Earth-like planet  \citep{held1994proposal}. This ideal climatology test focuses on the long-term and statistically-averaged final state. The final quasi-steady state of this test is similar to an Earth-like atmosphere with a latitudinal temperature gradient and large-scale winds. The original work applied Newtonian cooling and Rayleigh drag schemes to simplify the radiative forcing and boundary effects, respectively. The fluid motion is quasi-geostrophic under the equator-to-pole temperature gradient and the Coriolis force with Earth's rotational rate. In the long-term averaged equilibrium state, thermal wind theory predicts two sub-tropical jets.

The initial condition of our simulation is set as a quasi-hydrostatic atmosphere with random and small temperature perturbations deviated from dry adiabats to break the initial symmetry. The thermal structure is relaxed to a reference profile via a Newtonian cooling scheme,
\begin{equation}
    \frac{\partial E}{\partial t} + ... = - k_{T}(\phi, \sigma) \rho c_{v} [T - T_{eq}(\phi, p)] - \rho w g,
\end{equation}
where $c_{v}$ is the isochoric specific heat; $\sigma$ is the ratio of the local pressure, $p$, over the surface pressure, $p_s$; $k_{T}$ is the temperature damping strength as a function of the latitude and $\rm \sigma$; $T_{eq}$ is the reference temperature profile as a function of latitude and $\sigma$. $K_{T}$ and $T_{eq}$ are adopted from the original paper \citep{held1994proposal}.

Rayleigh drags are applied to all three momentum equations,
\begin{equation}
    \frac{\partial (\rho \boldsymbol{u})}{\partial t} + ... = - k_{v}(\sigma) \rho \boldsymbol{u} - 2\rho\boldsymbol{\Omega}\boldsymbol{\times}\boldsymbol{u} + \rho \boldsymbol{g}, 
\end{equation}
where $k_{v}$ is Rayleigh drag strength, $\boldsymbol{\Omega}$ is the planetary rotation vector. Geometric terms shall be written on the left-hand side of the equation. We treat both Newtonian cooling and Rayleigh drag friction as the explicit body forcing (i.e., explicit source terms), using the same parameters as used in the original paper \citep{held1994proposal}.

The simulation domain is 25 km in height, latitude from $\rm -90^\circ S$ to $\rm 90^\circ N$, and the longitude from $\rm 0^{\circ}$ to $\rm 360^{\circ}$. We adopt the spherical polar coordinate with the latitude and the longitude divided uniformly in degrees. Polar wedge scheme in $\rm Athena^{++}$ \citep{stone2020athena++} is applied to the polar region as the boundary conditions at the poles. Two solid-wall boundaries (i.e., reflecting or no flux boundary condition) are applied to both the bottom and the top of the domain (Table 1). The spatial resolution is about $\rm 2.8^{\circ}$ in latitude (64 cells), $\rm 2.8^{\circ}$ in longitude (128 cells), and 625 meters in height (40 layers). A 5-km-thick sponge layer (e.g., Rayleigh drag layer) is applied at the top of the atmosphere for both explicit and implicit schemes to absorb the vertical propagating acoustic and gravity waves. The CFL number for the explicit scheme is fixed as 0.3 which is calculated as CFL $ = c_{s} \Delta t/\Delta z$. We adopt CFL = 0.5 for the VIC scheme, in which case the CFL number is limited by the horizontal resolution, satisfying CFL $ = c_{s}\Delta t/\mathrm{min}(\Delta x,\Delta y)$.

Similar to the original work, the simulation reaches the quasi-equilibrium state after about 200 simulation days given the radiative cooling and Rayleigh drag timescales are about tens of days \citep{held1994proposal}. Simulation results are averaged from 200 days to 1200 days.

First, we present the averaged state of zonal-mean zonal wind, meridional wind, and temperature in Figure~\ref{fig:HS94-1}. We also present the eddy analysis of this test in Figure~\ref{fig:HS94-2} with the zonal-mean meridional eddy momentum flux and zonal-mean eddy kinetic energy. The maximum prograde jet speeds for explicit and implicit models are 30.97 $\rm m \; s^{-1}$ and 30.84 $\rm m \; s^{-1}$, respectively. The wind structures and jet speeds are very close to previously published simulations \citep[e.g.,][]{held1994proposal,lin2004vertically,ullrich2012mcore,mayne2013using, mendoncca2016thor}. 

\begin{figure}
    \centering
    \includegraphics[width=1.0\textwidth]{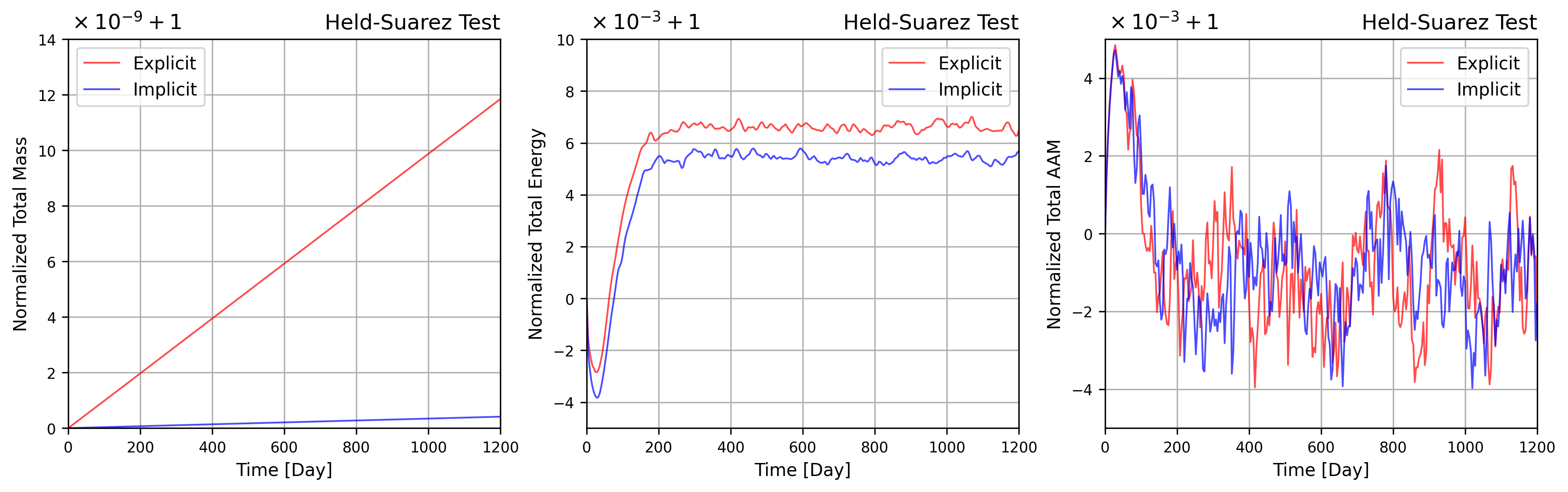}
    \caption{Temporal evolution of total mass, energy, and AAM normalized by the initial condition in the Held-Suarez atmospheric simulations from Day 0 to Day 1200. Explicit and implicit results are labeled with red and blue colors, respectively.}
    \label{fig:conservation-HS94}
\end{figure}

Second, in Figure~\ref{fig:conservation-HS94}, we show the temporal evolution of total mass, energy, and AAM of both the explicit and implicit simulation results. The fractional variation of total mass increases linearly at the machine precision level (i.e., $\rm \sim 10^{-9}$ of the initial value) at a rate of $\rm \sim 3\times 10^{-9}$ $\rm yr^{-1}$ and $\rm \sim 3\times 5^{-10}$ $\rm yr^{-1}$ for explicit and implicit results, respectively. The total mass changing behaviors are similar to the result of the Straka sinking bubble test, in which the VIC scheme results have less total mass change. It shows that the VIC scheme can conserve total mass just like the explicit scheme in a spherical polar coordinate system. The accumulated machine precision error of total mass is negligible given the total mass change is much smaller than $\rm 10^{-6}$ even the model with the VIC scheme is integrated for thousands of years. The total energy varies at the start by about 0.6\% and reaches a quasi-steady state after 200 days. The fractional variation of total energy is less than $10^{-3}$ after 200 days. The total AAM decreases from the initial value by about 0.1\% and varies by about 0.2\% after 200 days. Note that the total AAM is not expected to be rigorously conserved because we solve momentum equations instead of angular momentum equations. This result shows that our simulation results does achieve the steady after 200 days.

In sum, our model can reproduce the benchmark result for Earth-like global-scale atmospheric dynamics. The VIC scheme achieves the mass conservation to the machine precision level. In the steady state, the total energy and AAM vary at a sub-percent level.

Similar to previous tests, the VIC scheme can significantly improve computational efficiency. First, the time step increases by a factor of $\rm \sim$450 with the VIC scheme. Second, with using 512 CPU cores on Pleiades (Sandy Bridge processors), the computational efficiency is improved by more than two orders of magnitudes. The polar convergence issue of the spherical polar coordinate system still limits the spatial resolution in the polar region and therefore limits the time step and the computational efficiency, which can be improved by implementing a cubed-sphere coordinate system \citep{putman2007finite} or using Static Mesh Refinement (SMR) \citep{zhu2018global} in our future studies.

\subsection{Shallow Hot Jupiter Test} \label{sec:hj}

Our final global-scale test is on the atmospheres of hot Jupiters, which are Jovian-size extra-solar planets very close to their host stars. These giant planets are likely to be synchronously orbiting around their host stars due to strong gravitational tides, meaning the dayside of planets is always irradiated by their host stars. Even though most hot Jupiters could still be rapid rotators (rotational period of about three days) like Jupiter, the strong, permanent day-night irradiation patterns distinguish the climate state on these emerging planetary populations from any previously known planetary atmospheres in the Solar System. Observations and theories \citep[e.g.,][]{showman2002atmospheric,knutson20083,showman2009atmospheric} have shown that the equatorial super-rotating jet on hot Jupiters can be subsonic or sonic and the day-side and night-side temperature contrast can exceed more than 1000 K. Some ultra-hot Jupiters' atmospheres can be hotter than 2500 K so that hydrogen molecules on the dayside can be thermally dissociated \citep{bell2018increased,tan2019atmospheric}. To prepare our model for future application on diverse exoplanetary atmospheres, here we validate our global GCM in this new synchronously rotating hot Jupiter regime. In particular, we validate our model against an experiment called the shallow hot Jupiter test. Several previous models have performed this hot Jupiter benchmark test and showed relatively good agreement \citep{menou2009atmospheric,heng2011atmospheric,bending2012benchmark,mayne2014unified, mendoncca2016thor, mayne2017results}. But to date the published results were either from the model using primitive equations, hydrostatic models \citep{menou2009atmospheric,heng2011atmospheric,bending2012benchmark} or non-hydrostatic model with implicit and semi-implicit schemes \citep{mendoncca2016thor, mayne2017results}. In addition to testing our VIC scheme, here we also provide the first non-hydrostatic, explicit integration results for this case without damping the acoustic waves.

Similar to previous works, we simulate a canonical hot Jupiter with a planetary radius of $\rm 10^5$ km and the gravitational acceleration of 8 $\rm m \; s^{-2}$. The simulation domain covers 4500 km in height with 40 layers, $\rm -90^{\circ} S$ to $\rm 90^{\circ} N$ in latitude with 64 cells, and $\rm 0^{\circ}$ to $\rm 360^{\circ} $ in longitude with 128 cells. The bottom pressure and temperature are set as 1 bar and 1600 K, respectively. The initial temperature structure is isothermal (1600 K) from the surface to the top. The atmospheric temperature is relaxed to a reference temperature profile by a Newtonian cooling scheme, but no Rayleigh drag is applied. We adopt a 500 km thick sponge layer at the top of the domain to absorb vertically propagating waves. The same Newtonian cooling set-up is used as previous works. The reference temperature profile is given by

\begin{equation}
    T_{eq} = T_{vert} + \beta_{trop}\Delta T_{E-P} \cos{\lambda} \cos{\phi},
\end{equation}
where $\lambda$ is the longitude, $\phi$ is the latitude, $T_{vert}$ and $\beta_{trop}$ are parameters adopted from \cite{menou2009atmospheric}, \cite{heng2011atmospheric}, and \cite{mendoncca2016thor}. The equator-to-pole temperature difference, $T_{E-P}$, is 300 K. 

We choose the time steps corresponding to CFL = 0.3 and CFL = 0.9 for the explicit case and the implicit case, respectively.

\begin{figure}
    \centering
    \includegraphics[width=1.0\textwidth]{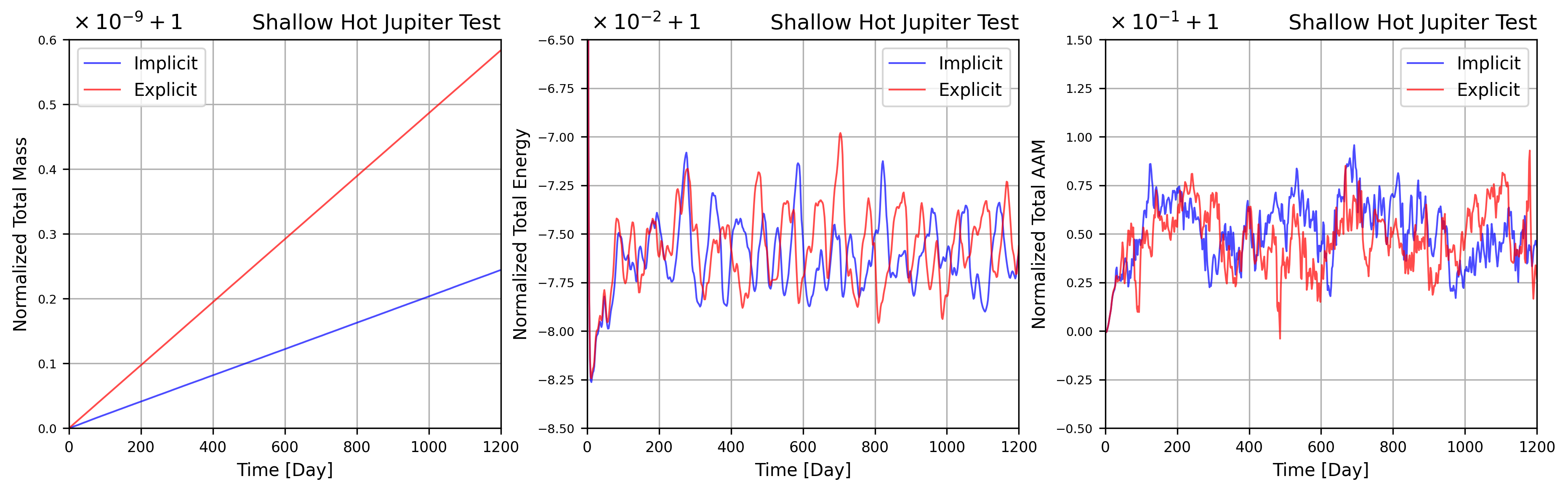}
    \caption{Temporal evolution of total mass, total energy, and total AAM normalized by the initial condition in shallow hot Jupiter simulations from Day 0 to Day 1200. Explicit and implicit results are labeled with red and blue colors, respectively.}
    \label{fig:HJ-conservation}
\end{figure}

\begin{figure}
    \centering
    \includegraphics[width=0.8\textwidth]{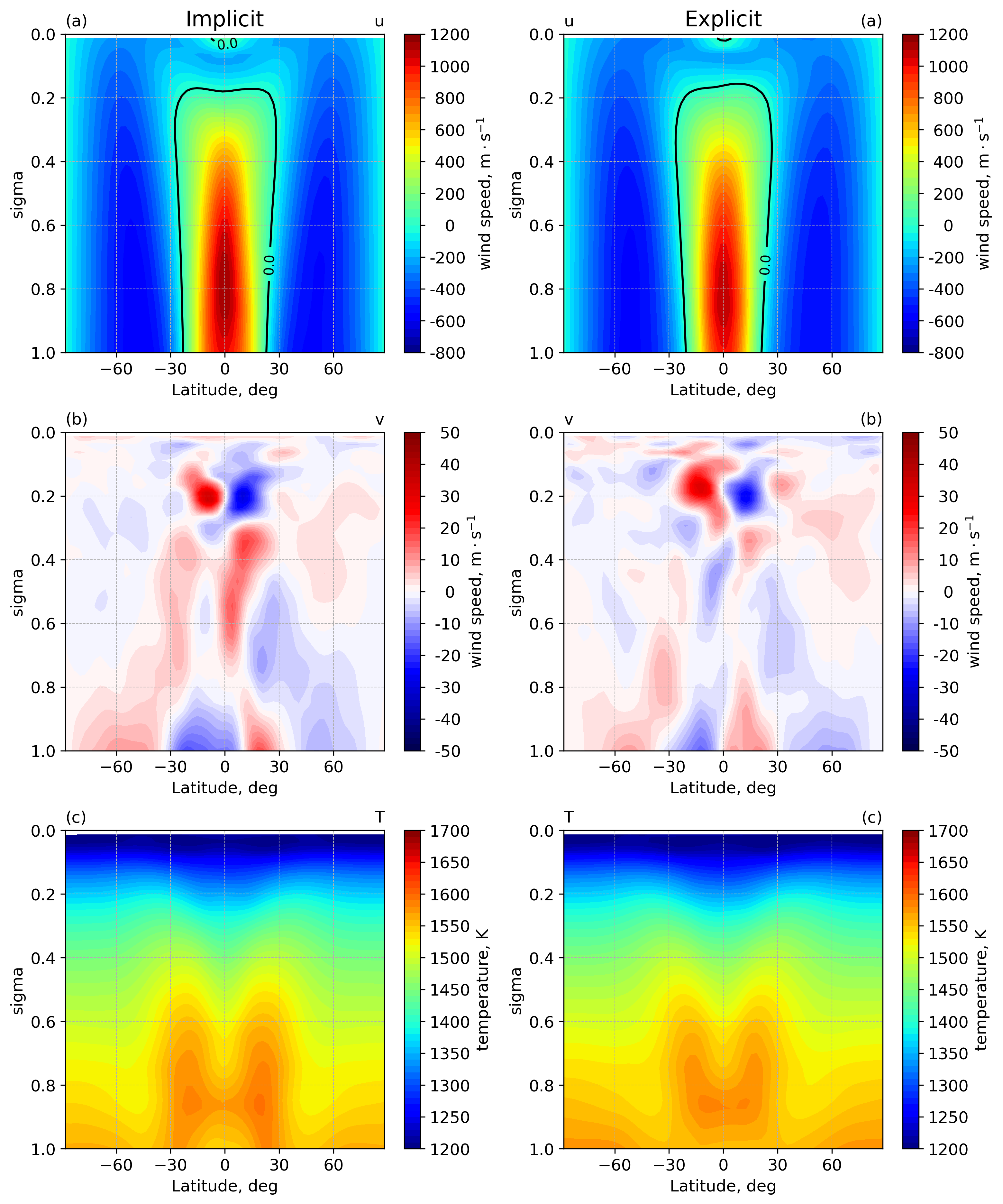}
    \caption{Explicit and implicit simulation results of shallow hot Jupiter test averaged over 1000 Earth days. Zonal-mean zonal wind, meridional wind, and temperature are shown in (a), (b), and (c), respectively.}
    \label{fig:HotJupiter1}
\end{figure}

\begin{figure}
    \centering
    \includegraphics[width=0.8\textwidth]{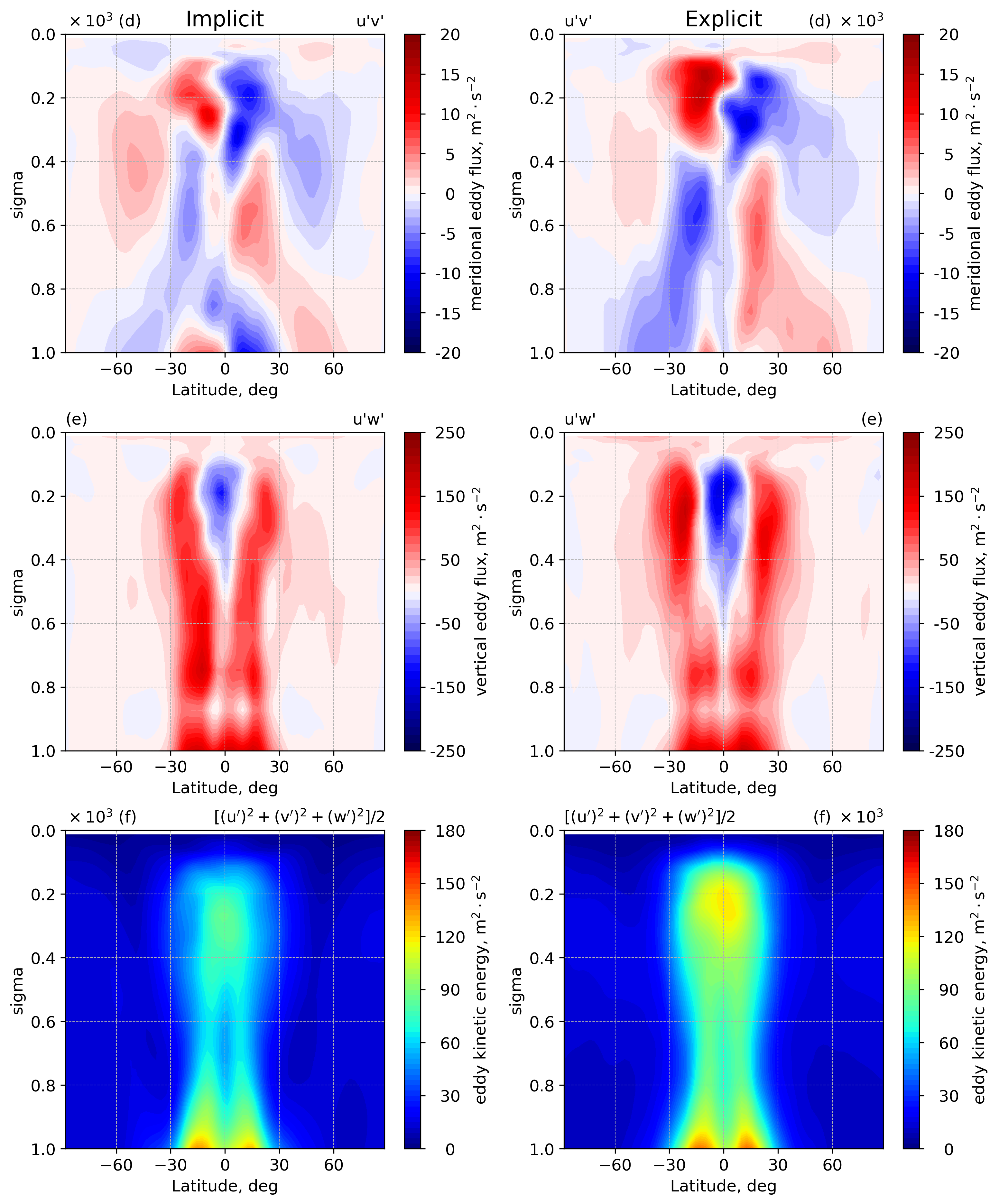}
    \caption{Explicit and implicit simulation results of shallow hot Jupiter test averaged for 1000 Earth days. Zonal-mean eddy meridional momentum flux, vertical eddy momentum flux, and eddy kinetic energy are plotted in (d), (e), and (f), respectively.}
    \label{fig:HotJupiter2}
\end{figure}

Figure~\ref{fig:HotJupiter1} shows the zonal-mean zonal wind, meridional wind, temperature structure in the statistically averaged state from Day 200 to Day 1200. Figure~\ref{fig:HotJupiter2} shows the zonal-mean meridional eddy momentum flux, vertical eddy momentum flux, and eddy kinetic energy. The zonal-mean zonal wind pattern exhibits a prograde equatorial super-rotating jet and retrogrades jets in the mid-latitude. The zonal-mean temperature pattern in the lower atmosphere shows a strong latitudinal variation, with two sub-tropical maxima and two polar maxima. The results from the explicit and VIC cases are almost identical. The maximum zonal-mean prograde zonal wind speeds are 1052.95 $\rm m \; s^{-1}$ and 1109.61 $\rm m \; s^{-1}$ for explicit and implicit simulations, respectively. The location of the super-rotating jet ranges from about $\rm -20^{\circ}S$ to $\rm 20^{\circ}N$. The jet speed difference is also associated with a slight difference in the equatorial temperature structure shown in Figure~\ref{fig:HotJupiter1}. The maximum westward jet speeds are -588.41 $\rm m \; s^{-1}$ and -581.59 $\rm m \; s^{-1}$ in the explicit and implicit cases, respectively. 

The zonal-mean zonal wind and zonal-mean temperature generally show similar structures to previous results. Both the explicit and VIC schemes produce a slightly slower equatorial jet than previous works \citep{menou2009atmospheric,heng2011atmospheric,bending2012benchmark,mayne2014unified, mendoncca2016thor, mayne2017results}. The zonal-mean temperature structure at the equator is also slightly different from previous models (note that these differences also exist among previous models). These subtle differences might come from different capabilities of resolving eddies in different models. Although the agreement between our implicit and explicit model results demonstrate a good performance of our VIC scheme in exoplanet simulations, future non-hydrostatic models with explicit time integration schemes would be needed to validate our explicit model results. 

It was also recently argued that simulations of the deep atmospheres of hot Jupiters might require a very long time integration to achieve the steady state \citep{mayne2017results,sainsbury2019idealised,deitrick2019thor,wang2020extremely}. The shallow hot Jupiter benchmark case has a relatively short radiative timescale and is expected to converge early. We also integrated our case for 12000 days to ensure that simulations have achieved the steady state but did not find a significant change after 1200 days. 

Figure~\ref{fig:HJ-conservation} shows the change of total mass, energy, and AAM for shallow hot Jupiter test in the first 1200 days. Both the explicit and implicit schemes show similar behavior (Figure~\ref{fig:HJ-conservation}). Similar to the Held-Suarez test, the total mass is well conserved at the machine-precision level with a slight, linear increase with time. In the steady state, total energy varies in a sub-percent level and total AAM varies within a few percents. Both the total energy and AAM at the steady state is different from their initial values by a few percents but shows no trend of continuous loss or increase, even after long-term integration of 12000 days.

\section{Conclusion} \label{sec:conclusion}

In this paper, we have developed a new non-hydrostatic planetary atmosphere model with a state-of-the-art vertically-implicit-correction scheme built on top of the $\rm Athena^{++}$ and SNAP framework. The VIC scheme has the advantage of satisfying conservation laws (i.e., conservation of total mass and energy for local simulations and conservation of total mass for global simulations). We validated the model using both localized simulations and global-scale simulations. Our VIC scheme can improve the computational efficiency of 3D global-scale simulations, especially for hydrogen-dominated atmospheres with the large aspect ratio of spatial resolutions between vertical direction and horizontal direction, which is relevant to Jupiter, Saturn, Uranus, Neptune, extra-solar gas giants, and brown dwarfs. The SNAP with the VIC scheme has several features:

1. With the VIC scheme, we are able to efficiently solve the full set of Euler equations on unstaggered grids by employing a dimensionally-unsplit method for atmospheric simulations. The VIC scheme significantly relaxes the CFL limitation and improves the computational efficiency compared with using the explicit scheme.

2. The algorithm of the VIC scheme in our model allows the use of different Riemann solvers (e.g., LMARS, Roe, HLLC), reconstruction methods (e.g., PPM, WENO5), and time integration schemes (e.g., RK3, VL2) in order to achieve different levels of spatial and temporal accuracy.

3. Linear wave test shows that the VIC scheme can use a large time step in atmospheric simulations by damping the amplitude of acoustic waves and suppress the numerical instability in time marching. The Straka sinking bubble test and the Robert rising bubble test show that, although the VIC scheme suppresses the spurious numerical noise of acoustic waves with numerical diffusion, it does not affect the ability of capturing the vertically-convective fluid motions even in the cases with weak forcing. These tests also show that our VIC scheme can well maintain numerical stability without any extra divergence damping or hyper-viscosity over the intrinsic damping in the VIC scheme. Users can also apply an additional eddy diffusion using the original diffusion solver in $\rm Athena^{++}$ if spatial resolution is not fine enough to capture the eddy transport.

4. The Earth-like Held-Suarez benchmark and shallow hot Jupiter test show that the VIC scheme performs well for 3D global simulations in the spherical polar coordinate system. The VIC scheme has been able to reproduce the numerical solutions of model validation tests published in the literature, showing that our VIC scheme canperform as the dynamical core for simulating atmospheric dynamics under different regimes. 

5. The analysis of the Straka sinking bubble, shallow hot Jupiter, and Held-Suarez tests shows that our VIC scheme has a good performance in conserving mass and energy in a closed domain and also maintaining the total AAM for global dynamic simulations. The fractional variation of total AAM is about 0.5\% and 10\% for Held-Suarez test and shallow Hot Jupiter test, respectively. The total AAM variation is commensurate with previous works \citep{mayne2017results, deitrick2019thor}. This capability provides a safeguard for simulations that requires long-time integration.

6. Both the local and the global tests show that the VIC scheme can significantly improve computational efficiency while quantitatively agreeing with the explicit results for large aspect ratio cases. For each time step, the VIC cases are generally only slower than the explicit cases by a factor of two to three. Because the VIC scheme allows a much larger time step, one can save the overall computational time in global-scale simulations by up to two orders of magnitude. In general, the VIC scheme can greatly reduce and save the simulation time and computational resources when applied to planetary atmospheric simulations.

We plan to improve the implicit SNAP in our future studies. Several improvements could include: coupling with a radiative transfer module, HARP, which was designed in the $\rm Athena^{++}$ framework \citep{li2018high}, developing the implicit cloud-resolving scheme to provide the capability of studying large-scale moist convection, implementing the cubed-sphere coordinate system to further improve the computational efficiency, and incorporating tracer transport modules with gas chemistry and cloud microphysics. Our ultimate goal is to develop a numerical scheme with topography so that we are able to study not only gas giants or aqua-planets but also study the atmospheric dynamics on terrestrial planets with realistic topography. Our implicit SNAP model will be made publicly available following the $\rm Athena^{++}$ open source policy.

\section{Acknowledgment}

We thank Zhaohuan Zhu, Chao-Chin Yang, Jo\~{a}o Mendon\c{c}a, Russell Deitrick, and Xiaoshan Huang for helpful discussions. We thank Nathan Mayne for important comments that have greatly improved this work. This research was supported by a NASA Earth and Space Science Fellowship 80NSSC18K1268 to H.G., 51 Pegasi b Fellowship to C.L., and NASA Solar System Workings grant NNX16AG08G to X.Z. We also acknowledge supercomputer Lux at UC Santa Cruz, funded by NSF MRI grant AST 1828315. We dedicate this work to Dr. Adam P. Showman (1968--2020) in recognition of his fundamental contribution to the understanding of atmospheric dynamics on giant planets and exoplanets.

\appendix
\section{Governing Equations in Different Coordinate Systems}
\label{sec:governing-equ}

The Euler equations in a Cartesian coordinate system can be written as
\begin{equation}
\label{equ:cart-continuity}
    \frac{\partial\rho}{\partial t}+\frac{\partial(\rho u)}{\partial x}+\frac{\partial (\rho v)}{\partial y}+\frac{\partial(\rho w)}{\partial z} = 0,
\end{equation}

\begin{equation}
\label{equ:cart-mom1}
    \frac{\partial(\rho u)}{\partial t} + \frac{\partial(\rho uu + p)}{\partial x} + \frac{\partial (\rho uv)}{\partial y} + \frac{\partial(\rho uw)}{\partial z} = 0,
\end{equation}

\begin{equation}
\label{equ:cart-mom2}
    \frac{\partial(\rho v)}{\partial t} + \frac{\partial(\rho uv)}{\partial x} + \frac{\partial (\rho vv+p)}{\partial y} + \frac{\partial(\rho vw)}{\partial z} = 0,
\end{equation}

\begin{equation}
\label{equ:cart-mom3}
    \frac{\partial(\rho w)}{\partial t} + \frac{\partial(\rho uw)}{\partial x} + \frac{\partial (\rho vw)}{\partial y} + \frac{\partial(\rho ww+p)}{\partial z} = -\rho g,
\end{equation}

\begin{equation}
\label{equ:cart-energy}
    \frac{\partial E}{\partial t}+\frac{\partial [u(E+p)]}{\partial x}+\frac{\partial [v(E+p)]}{\partial y}+\frac{\partial [w(E+p)]}{\partial z} = -\rho w g,
\end{equation}
where $g$ is the gravitational acceleration. The Euler equations set still adopt some approximations such as the spherically symmetric geopotential \citep[e.g.,][]{holton2004introduction,pedlosky2013geophysical,vallis2016geophysical,holton2016dynamic}.

For planetary-scaled simulations, the conservative form of the Euler equations should be treated in spherical coordinates, which can be written as

\begin{equation}
\label{equ:sphr-continuity}
    \frac{\partial\rho}{\partial t}+\frac{1}{r \cos{\theta}}\frac{\partial(\rho u)}{\partial \phi} + \frac{1}{r \cos{\theta}}\frac{\partial (\rho v \cos{\theta})}{\partial \theta} + \frac{1}{r^{2}} \frac{\partial (\rho r^{2} w)}{\partial r} = 0,
\end{equation}

\begin{equation}
\label{equ:sphr-mom1} 
    \frac{\partial(\rho u)}{\partial t} + \frac{1}{r \cos{\theta}}\frac{\partial(\rho uu)}{\partial \phi} + \frac{1}{r \cos{\theta}}\frac{\partial (\rho uv \cos{\theta})}{\partial \theta} + \frac{1}{r^{2}} \frac{\partial(\rho r^{2} uw)}{\partial r} = -\frac{1}{r \cos{\theta}} \frac{\partial p}{\partial \phi} + 
    \Big( 2\Omega+\frac{u}{r\cos{\theta}}\Big)(\rho v\sin{\theta}-\rho w\cos{\theta}),
\end{equation}

\begin{equation}
\label{equ:sphr-mom2}
    \frac{\partial(\rho v)}{\partial t} + \frac{1}{r \cos{\theta}}\frac{\partial(\rho uv)}{\partial \phi} + \frac{1}{r \cos{\theta}}\frac{\partial (\rho vv \cos{\theta})}{\partial \theta} + \frac{1}{r^{2}} \frac{\partial(\rho r^{2} vw)}{\partial r} = -\frac{1}{r} \frac{\partial p}{\partial \theta}
    -\frac{\rho vw}{r_{p}} - \Big( 2\Omega+\frac{u}{r\cos{\theta}}\Big)\rho u\sin{\theta},
\end{equation}

\begin{equation}
\label{equ:sphr-mom3}
    \frac{\partial(\rho w)}{\partial t} + \frac{1}{r \cos{\theta}}\frac{\partial(\rho uw)}{\partial \phi} + \frac{1}{r \cos{\theta}}\frac{\partial (\rho vw \cos{\theta})}{\partial \theta} + \frac{1}{r^{2}} \frac{\partial(\rho r^{2} ww)}{\partial r} = -\frac{\partial p}{\partial r}
    + \frac{\rho(u^2+v^2)}{r}+2\Omega\rho u\cos{\theta}-\rho g,
\end{equation}

\begin{equation}
\label{equ:sphr-energy}
    \frac{\partial E}{\partial t} + \frac{1}{r \cos{\theta}}\frac{\partial [u(E+p)]}{\partial \phi} + \frac{1}{r \cos{\theta}}\frac{\partial [v(E+p)\cos{\theta}]}{\partial \theta} + \frac{1}{r^{2}}\frac{\partial [w(E+p)]}{\partial r} = -\rho w g,
\end{equation}
where $r_{p}$ is planetary radius; $\Omega$ is the planetary rotational frequency; $\theta$ is latitude; $\phi$ is longitude; $r$ is the distance of the cell from the center of the planet.

\section{The Analytical Form of $\vert A \vert$}
\label{sec:explain-A}

The vector form of 1D Euler equations can be written as
\begin{equation}
    \frac{\partial \boldsymbol{Q}}{\partial t} + \frac{\partial \boldsymbol{F}}{\partial x} = 0.
\end{equation}

The quasi-linear form of the Euler equations can be written as
\begin{equation}
    \frac{\partial \boldsymbol{Q}}{\partial t} + \frac{\partial \boldsymbol{F}}{\partial \boldsymbol{Q}}\frac{\partial \boldsymbol{Q}}{\partial x} = 0.
\end{equation}
where ${\partial \boldsymbol{F}}/{\partial \boldsymbol{Q}}$ is the flux Jacobian. One can compute the analytical right eigenvectors $R$ and eigenvalues $\Lambda$,
\begin{equation}
    \label{equ:eigenvector}
    R =
    \begin{bmatrix}
    1                           & 1                       & 1            \\
    u-c_{s}                     & u                       & u+c_{s}      \\
    h-u c_{s}                   & h                       & h+u c_{s}    \\
    \end{bmatrix},
\end{equation}

\begin{equation}
    \label{equ:eigenvalue}
    \Lambda =
    \begin{bmatrix}
    u-c_{s}                     & 0                       & 0          \\
    0                           & u                       & 0          \\
    0                           & 0                       & u+c_{s}    \\
    \end{bmatrix},
\end{equation}
where $h = \gamma p/\rho(\gamma-1) + u^2/2$.

Then we can acquire $\vert A \vert$ by computing $R^{-1} \vert \Lambda \vert R$. Note that $\vert A \vert$ is automatically decided by the localized thermodynamic quantities (i.e., acoustic speed and flow speed).

\section{Conservation Laws under Reflecting Boundary Condition}
\label{sec:reflecting-prove}

Here, we prove that the conservation of total mass and energy is guaranteed in our VIC scheme with reflecting boundary conditions at both the top and bottom. We can write down the extended linear system with ghost cells as

\begin{equation}
\label{equ:reflecting-original}
    \begin{bmatrix}
    c_1 & a_1 + \frac{1}{\Delta t} & b_1 & 0 & \dots & 0 & 0 & 0 & 0 \\
    0 & c_2 & a_2 + \frac{1}{\Delta t} & b_2 & \dots & 0 & 0 & 0 & 0\\
    0& 0 & c_3 & a_3 + \frac{1}{\Delta t} & \dots & 0 & 0 & 0 & 0\\
    \dots & \dots & \dots & \dots & \dots & \dots & \dots & \dots & \dots \\
    0 & 0 & 0 & 0 & \dots & a_{m-2} + \frac{1}{\Delta t} & b_{m-2}& 0 & 0 \\
    0 & 0 & 0 & 0 & \dots & c_{m-1} & a_{m-1} + \frac{1}{\Delta t} & b_{m-1} & 0 \\
    0 & 0 & 0 & 0 & \dots & 0 & c_m & a_{m} + \frac{1}{\Delta t} & b_{m} \\
    \end{bmatrix}
    \begin{bmatrix}
    \Delta \boldsymbol{Q}_{0} \\ \Delta \boldsymbol{Q}_{1} \\ \Delta \boldsymbol{Q}_{2} \\ \Delta \boldsymbol{Q}_{3} \\ \dots \\ \Delta \boldsymbol{Q}_{m-2} \\ \Delta \boldsymbol{Q}_{m-1} \\ \Delta \boldsymbol{Q}_{m} \\ \Delta \boldsymbol{Q}_{m+1}
    \end{bmatrix}
    =
    \frac{1}{\Delta x}
    \begin{bmatrix}
    \boldsymbol{F}_{3/2}^{n} - \boldsymbol{F}_{1/2}^{n}\\
    \boldsymbol{F}_{5/2}^{n} - \boldsymbol{F}_{3/2}^{n}\\
    \boldsymbol{F}_{7/2}^{n} - \boldsymbol{F}_{5/2}^{n}\\
    \dots \\
    \boldsymbol{F}_{m-3/2}^{n} - \boldsymbol{F}_{m-5/2}^{n}\\
    \boldsymbol{F}_{m-1/2}^{n} - \boldsymbol{F}_{m-3/2}^{n}\\
    \boldsymbol{F}_{m+1/2}^{n} - \boldsymbol{F}_{m-1/2}^{n}\\
    \end{bmatrix},
\end{equation}
where $\boldsymbol{Q}_{0}^n$ and $\boldsymbol{Q}_{m+1}^n$ are ghost cells at the bottom and top boundaries. $\boldsymbol{Q}_{0}^n$ and $\boldsymbol{Q}_{m+1}^n$ are inferred by the first and the last cells in the domain, satisfying the boundary condition. They can be written as

\begin{equation}
    \label{equ:ghost}
    \boldsymbol{Q}_{0}^n
    =
    \begin{bmatrix}
    1  &  &  &  &  \\
    &  -1 &  &  &  \\
    &  &  1  &  &  \\
    &  &  &  1  &  \\
    &  &  &  &  1  \\
    \end{bmatrix} 
    \boldsymbol{Q}_{1}^n
    = M \boldsymbol{Q}_{1}^n
    ;
    \,
    \boldsymbol{Q}_{m+1}^n
    =
    \begin{bmatrix}
    1  &  &  &  &  \\
    &  -1 &  &  &  \\
    &  &  1  &  &  \\
    &  &  &  1  &  \\
    &  &  &  &  1  \\
    \end{bmatrix} 
    \boldsymbol{Q}_{m}^n
    = M \boldsymbol{Q}_{m}^n 
\end{equation}
where $M$ is the converting matrix. Then, we can substitute Equation~(\ref{equ:ghost}) into Equation~(\ref{equ:reflecting-original}) to simplify the matrix on the left-hand side of the linear equation. As a result, we can get an invertible tridiagonal matrix on the left-hand side,
\begin{equation}
\label{equ:reflecting-final}
    \begin{bmatrix}
    c_{1} M + a_1 + \frac{1}{\Delta t} & b_1 & 0 & \dots & 0 & 0 & 0 \\
    c_2 & a_2 + \frac{1}{\Delta t} & b_2 & \dots & 0 & 0 & 0 \\
    0 & c_3 & a_3 + \frac{1}{\Delta t} & \dots & 0 & 0 & 0 \\
    \dots & \dots & \dots & \dots & \dots & \dots & \dots \\
    0 & 0 & 0 & \dots & a_{m-2} + \frac{1}{\Delta t} & b_{m-2}& 0 \\
    0 & 0 & 0 & \dots & c_{m-1} & a_{m-1} + \frac{1}{\Delta t} & b_{m-1} \\
    0 & 0 & 0 & \dots & 0 & c_m & a_{m} + b_{m} M + \frac{1}{\Delta t} \\
    \end{bmatrix}
    \begin{bmatrix}
    \Delta \boldsymbol{Q}_{1} \\ \Delta \boldsymbol{Q}_{2} \\ \Delta \boldsymbol{Q}_{3} \\ \dots \\ \Delta \boldsymbol{Q}_{m-2} \\ \Delta \boldsymbol{Q}_{m-1} \\ \Delta \boldsymbol{Q}_{m} \\ 
    \end{bmatrix}
    =
    \frac{1}{\Delta x}
    \begin{bmatrix}
    \boldsymbol{F}_{3/2}^{n} - \boldsymbol{F}_{1/2}^{n}\\
    \boldsymbol{F}_{5/2}^{n} - \boldsymbol{F}_{3/2}^{n}\\
    \boldsymbol{F}_{7/2}^{n} - \boldsymbol{F}_{5/2}^{n}\\
    \dots \\
    \boldsymbol{F}_{m-3/2}^{n} - \boldsymbol{F}_{m-5/2}^{n}\\
    \boldsymbol{F}_{m-1/2}^{n} - \boldsymbol{F}_{m-3/2}^{n}\\
    \boldsymbol{F}_{m+1/2}^{n} - \boldsymbol{F}_{m-1/2}^{n}\\ 
    \end{bmatrix}
\end{equation}

Following the philosophy of Equation~(\ref{equ:compressed-linear}), we can acquire the change of total mass, momentum, and energy from

\begin{equation}
\label{equ:compressed-reflecting}
    \frac{1}{\Delta t} \sum_{i=1}^{m} \Delta \boldsymbol{Q}_i = 
    -(c_{1} M + a_1 + c_{2}) \Delta \boldsymbol{Q}_1 - (b_{m-1} + a_{m} + b_{m} M) \Delta \boldsymbol{Q}_m - \sum_{i=2}^{m-1} \big[ (b_{i-1} + a_{i} + c_{i+1}) \Delta \boldsymbol{Q}_i \big] + \frac{(\boldsymbol{F}_{m+1/2}^{n} - \boldsymbol{F}_{1/2}^{n})}{\Delta x}.
\end{equation}

As described in Section~\ref{sec:conservation-law}, the third term on Equation~(\ref{equ:compressed-reflecting})'s right-hand side is zero. The reflecting boundary condition also guarantees the mass and energy fluxes at boundaries are 0, $\boldsymbol{F}_{1/2}^{n} = (0, \Delta(\rho u)_{1/2}, 0)^{T}$ and $\boldsymbol{F}_{m+1/2}^{n} = (0, \Delta(\rho u)_{m+1/2}, 0)^{T}$. Therefore, the fourth term on the left-hand side is zero. The total mass and energy changes are only determined by the first and the second terms. We compute mass and energy changes in $(c_{1} M + a_1 + c_{2}) \Delta \boldsymbol{Q}_1$ and $(b_{m-1} + a_{m} + b_{m} M) \Delta \boldsymbol{Q}_m$ by analytically calculating $c_{1} M + a_1 + c_{2}$ and $b_{m-1} + a_{m} + b_{m} M$. By substituting Equation~(\ref{equ:a}), Equation~(\ref{equ:b}), and Equation~(\ref{equ:c}) into $c_{1} M + a_1 + c_{2}$ and $b_{m-1} + a_{m} + b_{m} M$, we can acquire

\begin{equation}
    \label{equ:appendix-1}
    c_{1} M + a_1 + c_{2} = \frac{1}{2}\Big[ (\vert A_{1/2}^n \vert - \vert A_{1/2}^n \vert M) - (\boldsymbol{J}_1^n + \boldsymbol{J}_{0}^n M) \Big]
\end{equation}
and
\begin{equation}
    \label{equ:appendix-2}
    b_{m-1} + a_{m} + b_{m} M = \frac{1}{2}\Big[ (\vert A_{n+1/2}^n \vert - \vert A_{n+1/2}^n \vert M) + (\boldsymbol{J}_m^n + \boldsymbol{J}_{m+1}^n M) \Big].
\end{equation}

Here, we just prove that the mass and energy changes in $(c_{1} M + a_1 + c_{2}) \Delta \boldsymbol{Q}_1$ are 0. A prove of Equation~(\ref{equ:appendix-2}) is similar to the prove of Equation~(\ref{equ:appendix-1}). $\boldsymbol{J}_{1}^n$ and $\boldsymbol{J}_{0}^n$ are determined by the density, velocity, and energy in the first cell's center, $\rho_1$, $u_1$, $E_1$,

\begin{equation}
    \label{equ:appendix-3}
    \boldsymbol{J}_1^n
    =
    \begin{bmatrix}
    0                              & 1                           & 0 \\
    (\gamma-3)\frac{u_1^2}{2}      & (3-\gamma)u_1                 & \gamma-1 \\
    -\big[(\gamma-2)\frac{u_1^2}{2}+\frac{\gamma}{\gamma-1}\frac{p_1}{\rho_1}\big]u_1 &
    \frac{\gamma}{\gamma-1}\frac{p_1}{\rho_1}+(3-\gamma)\frac{u_1^2}{2} &
    \gamma u_1 \\
    \end{bmatrix},
\end{equation}

\begin{equation}
    \label{equ:appendix-4}
    \boldsymbol{J}_{0}^n
    =
    \begin{bmatrix}
    0                              & 1                           & 0 \\
    (\gamma-3)\frac{u_1^2}{2}      & -(3-\gamma)u_1                 & \gamma-1 \\
    \big[(\gamma-2)\frac{u_1^2}{2}+\frac{\gamma}{\gamma-1}\frac{p_1}{\rho_1}\big]u_1 &
    \frac{\gamma}{\gamma-1}\frac{p_1}{\rho_1}+(3-\gamma)\frac{u_1^2}{2} &
    -\gamma u_1 \\
    \end{bmatrix},
\end{equation}
where $\vert A_{1/2}^n \vert$ is computed by $\rho_{1/2}$, $u_{1/2}$, and $E_{1/2}$ on cell interface between the first cell and the neighbouring ghost cell. Physically, the velocity at 1/2 cell interface should be 0. The numerical solver the vertical velocity at the boundary interface by the approximate Riemann solver (Roe scheme)

\begin{equation}
    \label{equ:appendix-5}
    u_{1/2} = \frac{\sqrt{\rho_{R}}u_{R}+\sqrt{\rho_{L}}u_{L}}{\sqrt{\rho_{R}}+\sqrt{\rho_{L}}},
\end{equation}
where the subscripts R and L represent the right and left states of the interface, respectively, which are acquired from the reconstructed density and velocity profiles in neighbouring cell centers. The reflecting boundary condition assures $\rho_R = \rho_L$ and $u_R = -u_L$. In this case, we can get the vertical velocity at the domain boundary, $u_{1/2} = 0$.

Then, we can simplify $\vert A_{1/2}^n \vert$ as

\begin{equation}
    \label{equ:appendix-6}
    \vert A_{1/2}^n \vert
    =
    \begin{bmatrix}
    1            & 1           & 1          \\
    -c_{s}       & 0           & c_{s}      \\
    H_{1/2}      & 0           & H_{1/2}    \\
    \end{bmatrix}
    \begin{bmatrix}
    c_{s}    &             &               \\
    &        & 0           &               \\
    &        &             & c_{s}         \\
    \end{bmatrix}
    \begin{bmatrix}
    0           & -\frac{1}{2c_{s}}  &  \frac{\gamma-1}{2c_{s}^2} \\
    1           & 0                  & -\frac{\gamma-1}{c_{s}^2} \\
    0           & \frac{1}{2c_{s}}   &  \frac{\gamma-1}{2c_{s}^2} \\
    \end{bmatrix}
    =
    \begin{bmatrix}
    0           & 0          & \frac{\gamma-1}{c_{s}}        \\
    0           & c_{s}      & 0                                 \\
    0           & 0          & \frac{\gamma-1}{c_{s}}H_{1/2} \\
    \end{bmatrix},
\end{equation}
where $c_s$ is the acoustic speed at cell interface $1/2$, $H$ is the total enthalpy, $H = \gamma p/(\gamma-1)+\rho u^2/2$. Then, we can substitute Equation~(\ref{equ:appendix-3}), (\ref{equ:appendix-4}), (\ref{equ:appendix-5}), and (\ref{equ:appendix-6}) into Equation~(\ref{equ:appendix-1}), we can get,

\begin{equation}
    \label{equ:appendix-7}
    c_{1} M + a_1 + c_{2} = 
    \frac{1}{2}\Big[ \vert A_{1/2}^n (\boldsymbol{\mathcal{I}} - M) - (\boldsymbol{J}_1^n + \boldsymbol{J}_{0}^n M) \Big] =
    \begin{bmatrix}
    0                           & 0                       & 0          \\
    (\gamma-3)\frac{u_{1}^2}{2} & c_{s}+(3-\gamma)u_1     & \gamma-1 \\
    0                           & 0                       & 0          \\
    \end{bmatrix}.
\end{equation}

Finally, we can compute the temporal mass and energy changes, $\Delta \rho_1$ and $\Delta E_1$, in $(c_{1} M + a_1 + c_{2}) \Delta \boldsymbol{Q}_1$ from this equation, which are 0,

\begin{equation}
    \frac{1}{\Delta t} \sum_{i=1}^{m} \Delta \boldsymbol{\rho}_i = 0; \, \frac{1}{\Delta t} \sum_{i=1}^{m} \Delta \boldsymbol{E}_i = 0
\end{equation}

\section{Conservation Laws under Double-Periodic Boundary Condition}
\label{sec:double-periodic-prove}

Here, we provide the supplement-detailed calculation on the conservation of total mass, momentum, and energy with double-periodic boundary conditions at the top and bottom. Equation~(\ref{equ:linear_sys}) can be specifically modified for double-periodic boundary conditions as

\begin{equation}
\label{equ:double-periodic}
    \begin{bmatrix}
    a_1 + \frac{1}{\Delta t} & b_1 & 0 & \dots & 0 & 0 & c_1 \\
    c_2 & a_2 + \frac{1}{\Delta t} & b_2 & \dots & 0 & 0 & 0 \\
    0 & c_3 & a_3 + \frac{1}{\Delta t} & \dots & 0 & 0 & 0 \\
    \dots & \dots & \dots & \dots & \dots & \dots & \dots \\
    0 & 0 & 0 & \dots & a_{m-2} + \frac{1}{\Delta t} & b_{m-2}& 0 \\
    0 & 0 & 0 & \dots & c_{m-1} & a_{m-1} + \frac{1}{\Delta t} & b_{m-1} \\
    b_m & 0 & 0 & \dots & 0 & c_m & a_{m} + \frac{1}{\Delta t} \\
    \end{bmatrix}
    \begin{bmatrix}
    \Delta \boldsymbol{Q}_{1} \\ \Delta \boldsymbol{Q}_{2} \\ \Delta \boldsymbol{Q}_{3} \\ \dots \\ 
    \Delta \boldsymbol{Q}_{m-2} \\ \Delta \boldsymbol{Q}_{m-1} \\ \Delta \boldsymbol{Q}_{m} 
    \end{bmatrix}
    =
    \frac{1}{\Delta x}
    \begin{bmatrix}
    \boldsymbol{F}_{3/2}^{n} - \boldsymbol{F}_{1/2}^{n}\\
    \boldsymbol{F}_{5/2}^{n} - \boldsymbol{F}_{3/2}^{n}\\
    \boldsymbol{F}_{7/2}^{n} - \boldsymbol{F}_{5/2}^{n}\\
    \dots \\
    \boldsymbol{F}_{m-3/2}^{n} - \boldsymbol{F}_{m-5/2}^{n}\\
    \boldsymbol{F}_{m-1/2}^{n} - \boldsymbol{F}_{m-3/2}^{n}\\
    \boldsymbol{F}_{m+1/2}^{n} - \boldsymbol{F}_{m-1/2}^{n}\\
    \end{bmatrix}
\end{equation}

For equation $i = 1$, the diagnostic variables in the ghost cell are inferred from the variables in the last cell in the domain. The same philosophy can be applied to $i = m$ equation. The total change of conserved variables, $\sum_{i=1}^{m}\Delta\boldsymbol{Q_{i}}$, can be acquired from Equation~(\ref{equ:double-periodic}),

\begin{equation}
\label{equ:compressed-periodic}
    \frac{1}{\Delta t} \sum_{i=1}^{m} \Delta \boldsymbol{Q}_i = 
    -(b_{m} + a_{1} + c_{2}) \Delta \boldsymbol{Q}_1 - (b_{m-1} + a_{m} + c_{1}) \Delta \boldsymbol{Q}_m - \sum_{i=2}^{m-1} \big[ (b_{i-1} + a_{i} + c_{i+1}) \Delta \boldsymbol{Q}_i \big] + \frac{(\boldsymbol{F}_{m+1/2}^{n} - \boldsymbol{F}_{1/2}^{n})}{\Delta x}.
\end{equation}

One can infer $\boldsymbol{F}_{m+1/2}^{n} = \boldsymbol{F}_{1/2}^{n}$, $b_{m} + a_{1} + c_{2} = 0$, and $b_{m-1} + a_{m} + c_{1} = 0$ from double-periodic boundary conditions. They assure the first, second, and fourth terms on the right-hand side of Equation~(\ref{equ:compressed-periodic}) to be 0. Then, we can acquire the conservation of total mass, momentum, and energy in a 1D double-periodic tube,

\begin{equation}
    \frac{1}{\Delta t} \sum_{i=1}^{m} \Delta \boldsymbol{Q}_i = 0.
\end{equation}

\end{document}